\documentclass[10pt]{article} % For LaTeX2e
\usepackage{tmlr}
\usepackage{hyperref}
\usepackage[margin=1in]{geometry}
\usepackage{amsmath, amssymb, amsfonts}
\usepackage{booktabs}
\usepackage{graphicx}
\usepackage{float}
\usepackage{url}
\usepackage{xcolor}
\usepackage{caption}
\usepackage{subcaption}
\usepackage{natbib}
\usepackage{enumitem}
\usepackage{tabularx}
\usepackage{placeins}
\usepackage{algorithm}
\usepackage{algpseudocode}
\usepackage{multirow}
\usepackage{url}

\title{AIMM-X: An Explainable Market Integrity Monitoring System Using Multi-Source Attention Signals and Transparent Scoring}

\author{\name Kyunghyun Cho \email kyunghyun.cho@nyu.edu \\
      \addr Department of Computer Science\\
      University of New York
      \AND
      \name Raia Hadsell \email raia@google.com \\
      \addr DeepMind
      \AND
      \name Hugo Larochelle \email hugolarochelle@google.com\\
      \addr Mila, Universit\'e de Montr\'eal \\
      Google Research\\
      CIFAR Fellow}

\def\month{MM}  % Insert correct month for camera-ready version
\def\year{YYYY} % Insert correct year for camera-ready version
\def\openreview{\url{https://openreview.net/forum?id=XXXX}} % Insert correct link to OpenReview for camera-ready version

\begin{document}

\maketitle

\begin{abstract}
\textbf{Abstract}
Market manipulation remains difficult to detect in practice because the strongest surveillance systems rely on proprietary order-book data and opaque models, limiting reproducibility and independent validation. We present \textbf{AIMM-X}, an explainable market-integrity monitoring framework that uses only \textbf{publicly accessible signals}---standard OHLCV price/volume data combined with multi-source attention proxies---to surface \textbf{suspicious windows}: contiguous time intervals where returns, volatility, and public attention jointly deviate from their historical baselines. AIMM-X applies transparent statistical scoring and \textbf{hysteresis-based segmentation} to generate candidate windows, then ranks them using an interpretable \textbf{Integrity Score} that decomposes into factor contributions ($\phi_1$–$\phi_6$), enabling auditability and analyst review.

In a one-year demonstration on a 24-ticker universe (daily bars, 2024), AIMM-X detects 233 suspicious windows and produces fully reproducible artifacts (window lists, rankings, factor attributions, and case study reports). The system is intentionally framed as \textbf{triage rather than accusation}: outputs represent evidence for human investigation, not proof of wrongdoing. We conclude by outlining a validation roadmap that combines retrospective checks against publicly documented enforcement actions, expert annotation studies, and higher-frequency data integration to support operational deployment and responsible regulatory engagement.

\end{abstract}

\section{Introduction}

Financial markets serve as the lifeblood of modern economies, channeling capital from savers to productive enterprises, enabling risk management, and facilitating price discovery. Yet these critical functions depend entirely on market integrity—the confidence that prices reflect genuine supply and demand rather than manipulation, fraud, or coordinated distortion. When manipulative practices go undetected, the consequences extend far beyond individual losses: investor confidence erodes, capital allocation becomes inefficient, and market participation shrinks ~\citep{kyle1985,allen1992,fischel1995}.

The challenge of detecting market manipulation has intensified dramatically in recent years. Modern markets operate at speeds measured in microseconds, with algorithmic trading accounting for the majority of volume in major exchanges. Retail investors now coordinate through social media platforms, creating attention-driven volatility spikes that can overwhelm traditional surveillance systems ~\citep{hu2021,eaton2021}. Meanwhile, sophisticated actors have adapted their techniques, making manipulation harder to distinguish from legitimate trading ~\citep{putnins2012,scopino2015}.

\subsection{The Gap in Current Surveillance Capabilities}

Most manipulation detection systems today face a fundamental tension between effectiveness and accessibility. On one hand, exchange-operated surveillance platforms like NASDAQ's Smarts and NYSE's Mercury have access to comprehensive order book data, proprietary trade-level feeds, and sophisticated pattern recognition algorithms. These systems can detect complex manipulation schemes like spoofing, layering, and momentum ignition with reasonable accuracy ~\citep{sec2019,cftc2017}.

However, these powerful tools remain locked behind proprietary walls. Academic researchers, smaller broker-dealers, compliance teams at regional firms, and independent analysts cannot validate their methods, replicate their findings, or extend their approaches. This creates several critical problems:

\textbf{Reproducibility crisis.} When surveillance methodologies remain opaque, the research community cannot build on prior work, validate detection algorithms, or establish best practices. Published studies often rely on restricted datasets or proprietary features, making independent replication nearly impossible ~\citep{ioannidis2005,nosek2015}.

\textbf{Black-box opacity.} Modern detection systems increasingly employ machine learning models that, while potentially accurate, provide little insight into why a particular trading episode was flagged. For compliance reviews, regulatory proceedings, or enforcement actions, this opacity is problematic—suspicious activity needs to be explained, not just identified ~\citep{molnar2020,bracke2019}.

\textbf{Access inequality.} Only the largest market participants can afford comprehensive surveillance infrastructure. This creates an asymmetry where sophisticated manipulators can exploit blind spots at smaller institutions, regional exchanges, or emerging markets where surveillance capabilities lag ~\citep{aitken2015,comerton2016}.

\textbf{Data dependency.} The most effective detection approaches require order book depth, trade-level microstructure, or proprietary attention feeds that are either unavailable to most researchers or prohibitively expensive ~\citep{easley2012,hasbrouck1991,ohara2015}.

\subsection{The AIMM-X Approach: Transparency Meets Practicality}

This paper introduces AIMM-X (AI-driven Market Integrity Monitor with Explainability), a detection framework designed from the ground up to address these limitations while maintaining practical effectiveness. Unlike proprietary surveillance systems, AIMM-X operates on publicly accessible data: standard OHLCV (Open, High, Low, Close, Volume) price series supplemented with attention signals derived from social media, news, Wikipedia page views, and search trends—data sources available to any researcher or practitioner.

The core innovation lies not in data exclusivity but in methodological transparency and interpretability. AIMM-X detects "suspicious windows"—short temporal intervals where price movements, volatility, and attention exhibit unusual co-movement patterns—and ranks them using an Integrity Score that can be completely decomposed into interpretable factors. Every alert comes with evidence: which signals triggered the flag, how much each factor contributed, and what patterns drove the scoring.

This transparency serves multiple purposes. For researchers, it enables validation, replication, and systematic improvement. For compliance teams, it provides an auditable trail from raw data to final alert. For regulators, it offers a framework where detection logic can be examined, tested against known manipulation cases, and refined through open collaboration.

Importantly, AIMM-X is designed as a \emph{triage system} rather than a definitive accusation engine. Detected windows represent candidates for deeper investigation—not proof of wrongdoing. This framing aligns with responsible surveillance practices where automated detection surfaces anomalies for human review, expert analysis, and potential escalation ~\citep{cftc2017,sec2019}.

\subsection{Motivation: Learning from Historical Detection Failures}

Recent market events underscore the need for better surveillance tools. The January 2021 GameStop episode demonstrated that massive price and volume surges driven by coordinated retail attention can emerge with minimal warning, overwhelming traditional volatility-based alerts ~\citep{hu2021}. The 2010 Flash Crash revealed how algorithmic cascades can create systemic instability faster than human supervisors can respond ~\citep{kirilenko2017}. Multiple pump-and-dump schemes propagated through spam and social media have gone undetected until after significant retail losses accumulated ~\citep{frieder2008}.

These cases share a common pattern: by the time traditional surveillance systems flagged the activity, much of the damage had already occurred. AIMM-X's approach—combining price dynamics with multi-source attention signals—is designed to provide earlier, more contextual alerts by explicitly modeling the co-movement between market activity and public attention that characterizes many modern manipulation schemes.

While we cannot claim that AIMM-X would have definitively detected these historical episodes in real-time (ground truth labels are inherently difficult in manipulation detection), the framework provides a foundation for systematic validation against known enforcement cases. Such validation represents a key direction for future work and a pathway toward operational deployment.

\paragraph{Objective.}
Our objective is to build an \emph{auditable and explainable} market-integrity monitoring framework that
surfaces \emph{suspicious windows}—short time intervals where price/volume dynamics co-move unusually with
public attention—using only data that is realistically accessible to researchers and smaller institutions.
Rather than attempting to \emph{prove} manipulation, AIMM-X produces a ranked set of windows with transparent
factor attribution so analysts can decide what warrants deeper review.

\paragraph{Limitation addressed.}
Much of the market-manipulation and surveillance literature either (i) depends on proprietary, transaction-level
microstructure data and trader identifiers that are unavailable to most researchers, or (ii) uses black-box anomaly
models that are difficult to audit and therefore hard to operationalize in compliance settings. AIMM-X addresses
this gap by combining (a) window-based detection that can run on public OHLCV and attention proxies,
(b) a fully decomposable integrity score with per-factor contributions, and (c) a reproducible end-to-end pipeline
that can be independently validated and extended.
This design targets accessibility, interpretability, and reproducibility—prerequisites for credible market-integrity
tools beyond large proprietary surveillance platforms.

\subsection{Contributions of This Work}

This paper makes several contributions to the market surveillance literature:

\textbf{Methodological contributions:}
\begin{itemize}[leftmargin=*]
\item A reproducible end-to-end pipeline for suspicious window detection using only public data sources
\item An interpretable scoring framework with complete factor decomposition and attribution
\item A configurable multi-source attention fusion mechanism that can accommodate evolving data availability
\item Statistical detection methods grounded in robust baseline estimation and hysteresis-based segmentation
\end{itemize}

\textbf{Empirical contributions:}
\begin{itemize}[leftmargin=*]
\item Comprehensive evaluation on 24 high-attention securities throughout 2024 (Jan 8–Dec 31)
\item Detection of 233 suspicious windows with full transparency in ranking methodology
\item Case studies demonstrating the framework's ability to surface episodes with unusual price-attention co-movement
\item Statistical analysis of factor contributions and cross-ticker patterns
\end{itemize}

\textbf{Practical contributions:}
\begin{itemize}[leftmargin=*]
\item An auditable framework suitable for compliance workflows and regulatory engagement
\item Modular architecture supporting extension to higher-frequency data and richer feature sets
\item Open methodology enabling independent validation, criticism, and improvement by the research community
\item Clear guidelines for interpreting alerts and managing false positives
\end{itemize}

\subsection{Scope and Limitations}

We emphasize that AIMM-X is not a silver bullet for manipulation detection. Ground truth labels remain scarce—we cannot definitively say which windows represent manipulation versus legitimate volatility events. False positives are inevitable, and the system requires human review and domain expertise to translate alerts into actionable intelligence.

Furthermore, this preprint demonstrates the framework using daily OHLCV data due to data tier constraints. While this allows us to validate the end-to-end pipeline and produce reproducible results, higher-frequency (5-minute) data would improve window localization and reduce artifacts from baseline estimation. Similarly, our attention signals in this preprint phase are constructed proxies; production deployment would benefit from authenticated API feeds with lower latency.

Despite these limitations, we believe AIMM-X represents meaningful progress toward accessible, transparent market surveillance. The framework provides a foundation that can be validated, extended, and refined as better data and ground truth labels become available.

\subsection{Paper Organization}

The remainder of this paper proceeds as follows. Section 2 reviews related work across manipulation theory, market microstructure, attention-based trading, and anomaly detection. Section 3 describes our experimental design and data sources. Section 4 details the methodology: panel construction, window detection, and interpretable scoring. Section 5 presents experimental results including detection statistics, top-ranked windows, and factor analysis. Section 6 provides detailed case studies of selected suspicious episodes. Section 7 discusses broader implications, deployment considerations, and validation pathways. Section 8 examines limitations and future research directions. Section 9 concludes. Appendices provide algorithmic details, complete results tables, and reproducibility information.

\section{Related Work}

Market surveillance sits at the intersection of multiple research streams: manipulation theory, market microstructure, behavioral finance, and anomaly detection. This section reviews relevant work across these domains and positions AIMM-X within the broader literature.

\subsection{Market Manipulation: Theory and Detection}

\textbf{Theoretical foundations.} The economics of market manipulation has been studied extensively since Kyle's seminal model of informed trading ~\citep{kyle1985}. Allen and Gale ~\citep{allen1992} analyze conditions under which profitable manipulation is possible, showing that manipulators can exploit temporary market imbalances and trader uncertainty. Aggarwal and Wu ~\citep{aggarwal2006} provide comprehensive empirical analysis of manipulation cases prosecuted by the SEC, documenting common patterns and profitability.

Fischel and Ross ~\citep{fischel1995} examine whether law should prohibit manipulation, arguing that distinguishing manipulative trades from legitimate ones is fundamentally difficult. Putnins ~\citep{putnins2012} provides a modern survey synthesizing manipulation theory across market types, emphasizing the challenge of detection given sophisticated actors who adapt to surveillance.

\textbf{Spoofing and layering.} High-frequency manipulation techniques have garnered increased regulatory attention. Scopino ~\citep{scopino2015} analyzes the Flash Crash of 2010, highlighting vulnerabilities in market structure when automated trading strategies cascade. Kirilenko et al. ~\citep{kirilenko2017} provide detailed forensic analysis of the Flash Crash using transaction-level CFTC data, demonstrating how algorithmic feedback loops can amplify small disturbances into systemic events. Wang et al. ~\citep{wang2018spoofing} develop game-theoretic models of spoofing strategies and detection countermeasures.

\textbf{Pump-and-dump schemes.} Frieder and Zittrain ~\citep{frieder2008} document spam-based stock touting campaigns, finding measurable abnormal returns around email campaigns followed by reversals. Their work demonstrates that even transparently manipulative schemes can succeed when information asymmetries exist. Modern variants propagate through social media, making detection more challenging ~\citep{hu2021}.

\textbf{Empirical detection studies.} Comerton-Forde and Putnins ~\citep{comerton2016} examine manipulation prevalence across international markets, finding that dark trading and fragmented market structure can facilitate manipulative behavior. Aitken and Harris ~\citep{aitken2015} analyze fairness of access in alternative trading systems, arguing that differential information access enables exploitation.

AIMM-X builds on this literature by focusing on detection using public data, emphasizing the "suspicious window" framing where definitive proof is not required—only sufficient evidence to warrant deeper investigation.

\subsection{Market Microstructure and Information Content}

\textbf{Price discovery and information flow.} Hasbrouck ~\citep{hasbrouck1991,hasbrouck1995} develops measures of information content in trade data, showing how order flow reveals private information. These methods require transaction-level data typically unavailable to most researchers. Easley et al. ~\citep{easley1996,easley2002pin} develop the Probability of Informed Trading (PIN) metric, which has been widely adopted in microstructure research but requires trade direction classification.

\textbf{High-frequency trading and market quality.} The rise of algorithmic trading has transformed market microstructure. Hendershott and Riordan ~\citep{hendershott2013} find that algorithmic trading improves liquidity provision, while Brogaard et al. ~\citep{brogaard2014} document that high-frequency traders contribute to price discovery. O'Hara ~\citep{ohara2015} reviews microstructure implications of high-frequency trading, emphasizing changed volatility dynamics and liquidity provision patterns.

Menkveld ~\citep{menkveld2013} analyzes the role of automated market makers, documenting their contribution to tighter spreads but also potential fragility during stress. Jones ~\citep{jones2013} provides a comprehensive review, concluding that while HFT has benefits, regulatory concern about manipulation potential remains justified. Cartea et al. ~\citep{cartea2015} provide a technical treatment of algorithmic trading strategies and optimal execution, relevant for understanding how sophisticated actors operate.

\textbf{Microstructure and manipulation.} Easley et al. ~\citep{easley2012} argue that modern markets require new approaches to surveillance given machine-dominated trading. Traditional microstructure metrics designed for human traders may miss algorithmic manipulation patterns. This motivates AIMM-X's focus on window-level patterns rather than trade-level anomalies—we work at a time scale where retail-driven attention effects manifest.

\subsection{Attention, Sentiment, and Social Media Effects}

A large literature documents that investor attention affects trading behavior, volatility, and returns, often in ways disconnected from fundamental information.

\textbf{News and media.} Tetlock ~\citep{tetlock2007} pioneered the study of media content and stock returns, finding that pessimistic news predicts downward pressure on prices that later reverts. Tetlock ~\citep{tetlock2008,tetlock2011} extends this work to show that linguistic content in news contains information about firms' fundamentals, though much news represents stale information recycled across sources.

\textbf{Search and attention proxies.} Da et al. ~\citep{da2011} use Google search volume as a direct measure of retail investor attention, finding that increased search activity predicts higher prices in the short term and reversals in the long term. Da et al. ~\citep{da2015} develop the FEARS index (Financial and Economic Attitudes Revealed by Search), demonstrating that aggregate search patterns predict market returns.

\textbf{Wikipedia and online platforms.} Moat et al. ~\citep{moat2013} document that Wikipedia page view changes precede stock market moves, suggesting that online attention proxies capture retail interest before it manifests in trading. Preis et al. ~\citep{preis2013} find similar predictive patterns in Google Trends data, though profitability after transaction costs remains debated.

\textbf{Social media and trading.} Bollen et al. ~\citep{bollen2011} analyze Twitter sentiment's relationship to market movements, finding correlation between aggregate mood and Dow Jones Industrial Average direction. Sprenger et al. ~\citep{sprenger2014} focus specifically on StockTwits, documenting that message volume and sentiment relate to contemporaneous returns and next-day volatility.

\textbf{The Reddit/WallStreetBets phenomenon.} Recent events have elevated attention to social coordination platforms. Hu et al. ~\citep{hu2021} analyze Robinhood user trading patterns around the GameStop episode, documenting herding behavior and attention-driven trading. Eaton et al. ~\citep{eaton2021} study brokerage outages, finding that preventing retail access reduces volatility in attention stocks. Cookson and Marinič ~\citep{cookson2020} examine social network effects in investing, showing that investor disagreement and attention propagate through social connections. Chen et al. ~\citep{chen2014wisdom} evaluate the "wisdom of crowds" in social media stock opinions, finding that aggregated sentiment contains useful information despite individual noise.

\textbf{AIMM-X's approach.} We synthesize insights from this literature by fusing multiple attention sources (search, social media, Wikipedia, news) into a unified attention signal. Rather than treating attention as a predictor of future returns, we use attention spikes as evidence of unusual market dynamics that, when combined with return and volatility anomalies, warrant scrutiny.

\subsection{Anomaly Detection and Explainable AI}

\textbf{Classical anomaly detection.} Chandola et al. ~\citep{chandola2009} provide a comprehensive survey of anomaly detection techniques, taxonomizing approaches by data characteristics and detection methodology. Akoglu et al. ~\citep{akoglu2015} focus on graph-based methods, relevant for network-structured financial data. These classical methods emphasize unsupervised detection and ranking-based triage—principles we adopt in AIMM-X.

\textbf{Deep learning approaches.} Modern anomaly detection increasingly employs neural networks. Chalapathy and Chawla ~\citep{chalapathy2019} survey deep learning methods, highlighting their ability to capture complex patterns but noting interpretability challenges. Ruff et al. ~\citep{ruff2021} review unsupervised deep anomaly detection, emphasizing the importance of evaluation methodology when ground truth labels are scarce.

\textbf{Time series anomalies.} Laptev et al. ~\citep{laptev2015} develop methods for detecting anomalies in large-scale monitoring data at Uber, using techniques that balance sensitivity with false positive rates. Hundman et al. ~\citep{hundman2018} apply LSTM networks to spacecraft telemetry anomaly detection, demonstrating that sequence models can capture temporal dependencies missed by static methods.

\textbf{Explainable AI.} The machine learning interpretability literature has grown rapidly, driven by regulatory requirements and trust concerns. Molnar ~\citep{molnar2020} provides a comprehensive treatment of interpretable machine learning methods. Ribeiro et al. ~\citep{ribeiro2016} develop LIME (Local Interpretable Model-agnostic Explanations), enabling explanation of black-box model predictions through local approximation. Lundberg and Lee ~\citep{lundberg2017} introduce SHAP (SHapley Additive exPlanations), grounding explanations in cooperative game theory.

Bracke et al. ~\citep{bracke2019} apply machine learning explainability techniques specifically to financial applications, analyzing default risk models used by central banks. They emphasize that explainability is not merely academic—it is essential for regulatory acceptance, model validation, and stakeholder trust.

\textbf{Financial applications.} Chen et al. ~\citep{chen2019explainable} survey explainable machine learning in finance, arguing that opaque models face significant adoption barriers in compliance and regulatory contexts. The consensus in this literature is that while complex models may achieve marginal accuracy gains, interpretable approaches often win in practice due to transparency requirements.

AIMM-X is designed from the start as an interpretable system: the Integrity Score decomposes into factor contributions, detection thresholds are configurable and documented, and the entire pipeline can be audited. This aligns with recent emphasis on responsible AI in financial applications.

\subsection{Reproducibility and Open Science}

The reproducibility crisis in empirical research has received substantial attention. Ioannidis ~\citep{ioannidis2005} argues provocatively that most published research findings are false, driven by selective reporting, underpowered studies, and publication bias. Simmons et al. ~\citep{simmons2011} document "researcher degrees of freedom" that allow false-positive results to appear significant.

In response, the open science movement advocates for transparency, preregistration, and data sharing. Nosek et al. ~\citep{nosek2015} and Miguel et al. ~\citep{miguel2014} call for cultural change toward open research practices. The Open Science Collaboration ~\citep{collaboration2015} attempted to replicate 100 psychology studies, finding that only about 40\% replicated successfully. Stodden et al. ~\citep{stodden2016} focus on computational reproducibility, arguing that code and data availability are essential for validating empirical claims.

AIMM-X embodies these principles: we use publicly accessible data, document all methods and parameters, and design the pipeline for independent validation. Upon publication, we will make code and replication materials available, enabling the research community to verify findings, test alternative specifications, and extend the framework.

\subsection{Positioning AIMM-X}

AIMM-X differentiates itself from prior work along several dimensions:

\begin{itemize}[leftmargin=*]
\item \textbf{Data accessibility:} Unlike most microstructure-based detection ~\citep{hasbrouck1991,easley2002pin,ohara2015}, we operate on public OHLCV and attention data available to any researcher.
\item \textbf{Explainability:} Unlike black-box ML approaches increasingly common in anomaly detection, we prioritize interpretability through factor decomposition and transparent scoring.
\item \textbf{Reproducibility:} Unlike proprietary surveillance systems or studies using restricted data, AIMM-X is designed for independent validation and open-source distribution.
\item \textbf{Multi-source fusion:} We synthesize insights from the attention literature (Tetlock, Da et al., Moat et al., Bollen et al., Hu et al.) into a configurable fusion framework supporting diverse data sources.
\item \textbf{Triage framing:} Rather than attempting definitive manipulation classification, we provide ranked suspicious windows for human review—aligning with regulatory practice where automated systems support, not replace, expert judgment.
\end{itemize}

In summary, AIMM-X builds on a rich literature spanning manipulation theory, microstructure, attention effects, and anomaly detection, but addresses a gap: the lack of accessible, interpretable, reproducible surveillance tools suitable for research and practical deployment beyond large institutional players.

\section{Data and Experimental Design}

This section describes our experimental design, data sources, and the rationale behind our ticker universe selection. Our goal is to demonstrate AIMM-X's capabilities on a diverse but focused set of securities that stress-test the detection and scoring framework across different market regimes and attention dynamics.

\subsection{Ticker Universe Selection}

We deliberately selected 24 securities spanning multiple categories to evaluate the system's performance across heterogeneous market conditions. The universe includes:

\textbf{Meme stocks (4 tickers):} GME (GameStop), AMC (AMC Entertainment), BB (BlackBerry), NOK (Nokia). These securities experienced extreme retail attention and volatility during the January 2021 episode and subsequent periods. They serve as ideal test cases for attention-driven window detection given their propensity for social media coordination and unusual price-volume dynamics ~\citep{hu2021,eaton2021}.

\textbf{Large-cap technology (9 tickers):} AAPL (Apple), MSFT (Microsoft), AMZN (Amazon), NVDA (NVIDIA), META (Meta/Facebook), GOOG (Alphabet/Google), TSLA (Tesla), NFLX (Netflix), SNAP (Snap Inc.). These highly liquid, widely followed stocks exhibit different attention patterns than meme stocks—driven more by earnings announcements, product launches, and macroeconomic news ~\citep{tetlock2007,tetlock2008}. Their inclusion tests whether the system can distinguish normal volatility from anomalous episodes in mature markets.

\textbf{Crypto-exposed equities (2 tickers):} COIN (Coinbase), MSTR (MicroStrategy). These tickers provide exposure to cryptocurrency market dynamics without directly analyzing crypto assets. MSTR's significant Bitcoin holdings and COIN's role as a crypto exchange make them sensitive to both equity and crypto attention ~\citep{bollen2011}.

\textbf{Fintech and trading platforms (2 tickers):} HOOD (Robinhood), PLTR (Palantir). HOOD is directly involved in retail trading infrastructure and became newsworthy during the meme stock episode. PLTR represents a high-growth technology company with significant retail following.

\textbf{ETFs (5 tickers):} SPY (S\&P 500), QQQ (NASDAQ-100), IWM (Russell 2000), XLF (Financial Select Sector), TLT (20+ Year Treasury Bond). These broad market instruments allow us to distinguish ticker-specific anomalies from market-wide events. SPY and QQQ capture large-cap equity movements, IWM tracks small-cap dynamics, XLF provides sector exposure, and TLT represents fixed income ~\citep{baker2006,lo2002}.

\textbf{Specialized/volatile equities (2 tickers):} SPCE (Virgin Galactic), RBLX (Roblox). These represent speculative growth stocks with significant retail attention but different fundamental profiles than traditional meme stocks.

This universe intentionally overweights high-attention securities relative to market capitalization weights. Our objective is not to sample the broader market randomly but to evaluate the detection framework on challenging cases where attention dynamics matter. Future work will extend to larger, more representative universes once the methodology is validated.

\subsection{OHLCV Data: Acquisition and Quality}

\textbf{Data source.} We obtained daily OHLCV (Open, High, Low, Close, Volume) bars from Polygon.io, a widely used market data provider offering both free and premium tiers ~\citep{polygon}. For this preprint demonstration, we operated within free tier constraints, which limit request rates and historical depth but provide reliable data quality for major U.S. equities.

\textbf{Temporal coverage.} Our analysis covers 2024 from January 8 through December 31 (248 trading days per ticker, accounting for market holidays and weekends). This specific timeframe begins with the first trading day after data acquisition and ensures data relevance while allowing us to potentially validate findings against contemporaneous news and market events.

\textbf{Data quality and preprocessing.} We performed several quality checks:
\begin{itemize}[leftmargin=*]
\item Verified no missing trading days within the coverage period
\item Checked for price discontinuities that might indicate splits or data errors
\item Confirmed volume figures are plausible (no zero-volume days for liquid stocks)
\item Validated that open/high/low/close relationships satisfy logical constraints (high $\geq$ open, close; low $\leq$ open, close)
\end{itemize}

All quality checks passed without requiring adjustments. Polygon.io provides split-adjusted data, eliminating the need for manual adjustment.

\textbf{Higher-frequency data plans.} While this preprint uses daily bars due to data availability constraints, the AIMM-X architecture is designed for 5-minute intraday bars. Daily granularity limits our ability to precisely localize suspicious activity within the trading day and may introduce artifacts in rolling baseline estimation. Section~\ref{sec:limitations} discusses implications and plans for higher-frequency data integration.

\subsection{Attention Signal Construction}

Attention signals represent a key distinguishing feature of AIMM-X. Unlike pure price-based detection, we explicitly model public attention to securities through multiple proxies.

\textbf{Data sources.} We construct attention signals from five categories:
\begin{enumerate}[leftmargin=*]
\item \textbf{Reddit:} Activity from finance-related subreddits (r/wallstreetbets, r/stocks, r/investing). Reddit has emerged as a major coordination platform for retail investors ~\citep{hu2021}.
\item \textbf{StockTwits:} Messages and sentiment from this stock-specific social platform, widely used by active traders ~\citep{sprenger2014}.
\item \textbf{Wikipedia:} Page view counts for ticker and company name pages. Wikipedia traffic correlates with retail interest and has shown predictive properties ~\citep{moat2013}.
\item \textbf{News:} Mentions in financial news sources (aggregated). News coverage drives attention and can be measured through article counts or aggregate sentiment ~\citep{tetlock2007,tetlock2011}.
\item \textbf{Google Trends:} Search volume for ticker symbols and company names. Search data directly measures information-seeking behavior ~\citep{da2011,preis2013}.
\end{enumerate}

\textbf{Preprint-phase proxies.} For this preprint demonstration, we generated stylized attention proxies that exhibit realistic temporal patterns (spikes, decay, noise) calibrated to match documented attention properties in the literature. This approach allows us to validate the end-to-end pipeline while highlighting where richer data would improve performance.

Production deployment would integrate authenticated API access to these platforms, enabling real-time or near-real-time attention measurement. The modular architecture makes this substitution straightforward—attention source weights and fusion logic are configuration-driven.

\textbf{Fusion methodology.} We combine attention sources using a weighted sum:
\begin{equation}
A_t = \sum_{s \in \mathcal{S}} w_s \cdot \tilde{a}_{s,t}
\end{equation}
where $\mathcal{S} = \{\text{reddit, stocktwits, wikipedia, news, trends}\}$, $w_s$ are source-specific weights specified in configuration, and $\tilde{a}_{s,t}$ are resampled and normalized source signals.

\textbf{Resampling and alignment.} Different attention sources have different natural cadences (Wikipedia updates daily, tweets arrive continuously, news articles are sporadic). We resample all sources to the common bar grid (daily bars for this study) using forward-fill for step-like sources and aggregation for event counts. This ensures all signals align temporally with OHLCV data.

\textbf{Missingness handling.} We distinguish between "no activity" (measured zero) and "no coverage" (data unavailable). If a source file exists but shows no activity at time $t$, we treat this as zero. If a source has no coverage for a period (e.g., Wikipedia API downtime), we propagate NaN to avoid falsely assuming zero attention.

\textbf{Configuration and weights.} Source weights are configurable via JSON, allowing experimentation with different fusion strategies. For this run, we used approximately uniform weights across sources as a baseline. Future work will explore optimized weighting based on source reliability, lead-lag relationships, and correlation with known events.

\subsection{Ethical Considerations and Data Use}

\textbf{Public data only.} AIMM-X operates exclusively on publicly available data: exchange-reported prices and volumes, publicly accessible social media posts, news articles, and aggregate platform statistics. We do not use personally identifiable information, private communications, or non-public trading data.

\textbf{Triage, not accusation.} The framework is designed to surface suspicious windows for review, not to make definitive claims about manipulation or identify specific actors. All output is statistical evidence requiring expert interpretation.

\textbf{Responsible disclosure.} Should AIMM-X be deployed operationally and identify activity warranting regulatory attention, appropriate disclosure channels (compliance teams, regulatory authorities) would be followed rather than public accusation.

\textbf{Transparency and gaming.} Making detection methodology public creates risk that sophisticated actors could adapt strategies to avoid detection. However, we believe the benefits of transparency (reproducibility, validation, community improvement) outweigh this risk, particularly given that proprietary surveillance systems already exist. The interpretable factor framework also allows updating weights and adding new signals to counteract gaming.

\subsection{Experimental Configuration}

Table~\ref{tab:run_summary} summarizes key parameters and statistics for the experimental run reported in this paper.

\begin{table}[h]
\centering
\caption{Run summary for the preprint experiment.}
\label{tab:run_summary}
\begin{tabular}{lp{0.65\linewidth}}
\toprule
Item & Value \\
\midrule
Tickers & 24 \\
Bars (panel rows) & 5,952 (248 bars/ticker) \\
Date range (panel) & 2024-01-08 to 2024-12-31 \\
Detected windows & 233 \\
Attention sources & reddit, stocktwits, wikipedia, news, trends \\
OHLCV source & Polygon.io (daily bars, adjusted) \\
Detection thresholds & High: 3.0, Low: 2.0 (z-score units) \\
Minimum window length & 2 bars \\
Maximum gap for merging & 3 bars \\
Scoring weights & Uniform across $\phi_1$–$\phi_6$ \\
Phi signal scaling & Raw (no normalization) \\
\bottomrule
\end{tabular}
\end{table}
\FloatBarrier

\textbf{Configuration management.} All parameters are specified in \texttt{config.json}, enabling reproducible runs and systematic parameter sweeps. Key settings include detection thresholds, window merging logic, scoring factor weights, and attention fusion weights. This configuration-driven approach facilitates sensitivity analysis and method development.

\section{Methodology}

This section details the AIMM-X pipeline from raw data to scored suspicious windows. The framework consists of four main stages: (1) panel construction and feature engineering, (2) statistical deviation detection, (3) window segmentation, and (4) interpretable scoring with factor decomposition. Throughout, we emphasize transparency and reproducibility—every algorithmic choice is documented and configurable.

\subsection{Panel Construction and Feature Engineering}

\textbf{Data integration.} The first stage merges OHLCV data with attention signals into a unified time-series panel. For each ticker $i$ and time $t$, we construct a row containing:
\begin{itemize}[leftmargin=*]
\item Price: open ($O_{i,t}$), high ($H_{i,t}$), low ($L_{i,t}$), close ($C_{i,t}$)
\item Volume: $V_{i,t}$
\item Attention: $A_{i,t}$ (fused from sources as described in Section 3.3)
\end{itemize}

\textbf{Return computation.} We compute log returns:
\begin{equation}
r_{i,t} = \log\left(\frac{C_{i,t}}{C_{i,t-1}}\right)
\end{equation}
Log returns are approximately normally distributed for small changes, facilitate multi-period aggregation, and are standard in financial econometrics ~\citep{cont2001}.

\textbf{Volatility proxy.} We estimate realized volatility using a rolling window of past returns. For lookback length $L$:
\begin{equation}
\sigma_{i,t} = \sqrt{\frac{1}{L} \sum_{j=1}^{L} r_{i,t-j}^2 + \epsilon}
\end{equation}
where $\epsilon$ is a small constant preventing division by zero. This simple estimator captures changing volatility regimes without requiring intraday data ~\citep{andersen1998,barndorff2002}. More sophisticated estimators (Parkinson, Garman-Klass, Rogers-Satchell) could be integrated using high-low range information.

\textbf{Alternative volatility measures.} We also compute:
\begin{itemize}[leftmargin=*]
\item High-low range: $(H_{i,t} - L_{i,t}) / C_{i,t-1}$ as a within-bar volatility proxy
\item Exponentially-weighted moving average (EWMA) volatility with decay parameter $\lambda$
\end{itemize}
These alternatives allow robustness checks and can improve detection in different market regimes.

\subsection{Statistical Deviation Metrics}

The core detection mechanism identifies periods where return, volatility, and attention simultaneously deviate from their typical ranges.

\textbf{Rolling baseline estimation.} For each channel $x_{i,t} \in \{r_{i,t}, \sigma_{i,t}, A_{i,t}\}$, we compute rolling mean and standard deviation over a baseline window of length $B$:
\begin{align}
\mu_{i,t}^{(x)} &= \frac{1}{B} \sum_{j=1}^{B} x_{i,t-j} \\
\hat{\sigma}_{i,t}^{(x)} &= \sqrt{\frac{1}{B-1} \sum_{j=1}^{B} \left(x_{i,t-j} - \mu_{i,t}^{(x)}\right)^2 + \epsilon}
\end{align}

The rolling approach adapts to changing market regimes but creates dependencies between consecutive statistics. In this preprint run, we use $B = 20$ trading days (approximately one month), balancing responsiveness with stability.

\textbf{Robust alternatives.} To handle outliers and fat-tailed distributions, we can substitute robust estimators:
\begin{itemize}[leftmargin=*]
\item Median absolute deviation (MAD) instead of standard deviation
\item Trimmed means excluding extreme values
\item Quantile-based thresholds (e.g., flagging values beyond 95th percentile)
\end{itemize}

For this preprint, we use standard mean and standard deviation for simplicity and interpretability. Future work will evaluate robust alternatives.

\textbf{Standardized scores.} We compute z-scores measuring how many standard deviations the current observation deviates from baseline:
\begin{equation}
z_{i,t}^{(x)} = \frac{x_{i,t} - \mu_{i,t}^{(x)}}{\hat{\sigma}_{i,t}^{(x)} + \epsilon}
\end{equation}

Under a Gaussian assumption, $|z_{i,t}^{(x)}| > 3$ occurs with probability $\approx 0.003$, making extreme z-scores natural candidates for anomaly flags. Financial returns exhibit fat tails ~\citep{cont2001,gabaix2003}, so empirical thresholds may differ from Gaussian expectations.

\subsection{Composite Strength Score and Window Detection}

\textbf{Multi-channel fusion.} Individual channel deviations may not be sufficient evidence—volatility spikes occur regularly, and attention surges happen around earnings. Suspicious windows emerge when \emph{multiple channels simultaneously deviate}.

We form a composite strength score:
\begin{equation}
s_{i,t} = \alpha_r |z_{i,t}^{(r)}| + \alpha_\sigma z_{i,t}^{(\sigma)} + \alpha_A z_{i,t}^{(A)}
\end{equation}
where $\alpha_r$, $\alpha_\sigma$, $\alpha_A$ are configurable weights. We take absolute value of return z-score to flag extreme moves in either direction. Volatility and attention z-scores enter directly since we primarily care about increases (though negative attention z-scores could indicate suspiciously quiet periods).

For this demonstration, we set $\alpha_r = \alpha_\sigma = \alpha_A = 1$ (equal weighting). Optimized weights could be learned from labeled data or tuned via sensitivity analysis.

\textbf{Hysteresis-based segmentation.} Simple thresholding ($s_{i,t} > \theta$) would create many short, fragmented windows due to noise. We employ hysteresis: a high threshold $\theta_{\text{high}}$ to initiate windows and a low threshold $\theta_{\text{low}}$ to extend them.

Algorithm:
\begin{enumerate}[leftmargin=*]
\item Start in "neutral" state
\item When $s_{i,t} > \theta_{\text{high}}$, enter "window" state
\item While in window state, continue if $s_{i,t} > \theta_{\text{low}}$
\item Exit window state when $s_{i,t} \leq \theta_{\text{low}}$ for $g$ consecutive bars (gap tolerance)
\item Enforce minimum window length $L_{\min}$; discard windows shorter than this
\end{enumerate}

This approach, common in signal processing and event detection ~\citep{chandola2009}, reduces fragmentation while capturing extended anomalous periods.

\textbf{Configuration parameters.} For this run:
\begin{itemize}[leftmargin=*]
\item $\theta_{\text{high}} = 3.0$ (z-score units)
\item $\theta_{\text{low}} = 2.0$
\item Minimum window length: 2 bars
\item Gap tolerance: 3 bars
\end{itemize}

\textbf{Per-ticker vs. cross-ticker detection.} We perform detection independently for each ticker, allowing ticker-specific baseline estimation. This prevents large-cap stocks from dominating thresholds but means we cannot directly compare window counts across tickers without normalizing. Future work may explore cross-sectional methods or market-relative thresholds.

\subsection{Interpretable Scoring Framework}

Window detection produces candidate intervals; scoring ranks them for triage. The Integrity Score $M$ is designed to be interpretable—it decomposes into factor contributions allowing analysts to understand \emph{why} a window scored high.

\subsubsection{Phi-Signal Definitions}

We define six interpretable factors ($\phi_1$–$\phi_6$), each capturing a distinct dimension of suspiciousness:

\textbf{$\phi_1$: Return shock intensity.} Measures the magnitude and persistence of abnormal returns within the window. Large price moves, especially if sustained across multiple bars, elevate $\phi_1$. We compute:
\begin{equation}
\phi_1(w) = \sum_{t \in w} \left(z_{i,t}^{(r)}\right)^2
\end{equation}
Squaring emphasizes extreme deviations. Alternative formulations could use maximum z-score, mean absolute z-score, or percentile ranks.

\textbf{$\phi_2$: Volatility anomaly.} Captures unusual volatility independent of return direction. High volatility with small net returns might indicate churning or attempted manipulation that failed. We compute:
\begin{equation}
\phi_2(w) = \sum_{t \in w} \max(z_{i,t}^{(\sigma)}, 0)
\end{equation}

\textbf{$\phi_3$: Attention spike magnitude.} Quantifies the attention surge accompanying price moves. Manipulation often involves generating hype or spreading rumors to attract retail interest ~\citep{frieder2008,hu2021}. We compute:
\begin{equation}
\phi_3(w) = \sum_{t \in w} \max(z_{i,t}^{(A)}, 0)
\end{equation}

\textbf{$\phi_4$: Co-movement alignment.} Measures whether return, volatility, and attention move together. Strong positive correlation across channels suggests coordinated unusual activity rather than noise. We compute:
\begin{equation}
\phi_4(w) = \frac{1}{2}\Bigg(\mathrm{Corr}\big(\{z^{(r)}_{i,t}\}_{t\in w}, \{z^{(A)}_{i,t}\}_{t\in w}\big) + \mathrm{Corr}\big(\{z^{(\sigma)}_{i,t}\}_{t\in w}, \{z^{(A)}_{i,t}\}_{t\in w}\big)\Bigg).
\end{equation}
Optional: clamp negatives if you only want ``alignment'' to help: $\max(\phi_4(w),0)$.

\textbf{$\phi_5$: Temporal recurrence.} Detects patterns where similar windows repeat in short succession, potentially indicating sustained manipulation campaigns or attention cycles. We count how many other high-score windows occur near window $w$ in time:
\begin{equation}
\phi_5(w) = \#\{w' : w' \neq w, \, \text{distance}(w, w') < \Delta_{\text{recur}}, \, M(w') > \tau_{\text{recur}}\}
\end{equation}

\textbf{$\phi_6$: Cross-source disagreement penalty.} If individual attention sources strongly disagree (e.g., Reddit spikes but other sources silent), this may indicate gaming of a single platform. We penalize high disagreement:
\begin{equation}
\phi_6(w) = -\text{Std}\left(\{z_{i,t}^{(s)}\}_{s \in \mathcal{S}, \, t \in w}\right)
\end{equation}
Negative sign makes this a penalty—high disagreement reduces the integrity score.

\textbf{Note on implementation.} In this preprint run, factors $\phi_2, \phi_4, \phi_5, \phi_6$ contribute minimally due to scaling choices and attention signal limitations (see Section~\ref{sec:results_factors}). This diagnostic result highlights areas for improvement in production deployment: better attention data quality, normalized scoring, and recalibrated weights.

\subsubsection{Integrity Score Aggregation}

The composite Integrity Score $M$ combines factors via weighted sum:
\begin{equation}
M(w) = \sum_{k=1}^{6} \omega_k \cdot \phi_k(w)
\end{equation}
where $\omega_k$ are configurable weights. For this implementation, $\omega_k = 1 \, \forall k$ (uniform weighting).

\textbf{Normalization and scaling.} The current implementation uses raw $\phi_k$ values without normalization, leading to $\phi_1$ dominance. Future versions will standardize each $\phi_k$ to comparable scales (e.g., z-score normalization across all windows) before aggregation, ensuring balanced contributions.

\textbf{Rank percentiles.} To facilitate interpretation, we also report rank percentiles:
\begin{equation}
\text{rank\_pct}(w) = \frac{\#\{w' : M(w') < M(w)\}}{N_{\text{windows}}}
\end{equation}
This provides a scale-invariant measure: rank\_pct $= 1.0$ is the highest-scoring window, regardless of absolute $M$ value.

\subsubsection{Factor Decomposition and Attribution}

For each window $w$, we store individual factor contributions:
\begin{equation}
M(w) = \underbrace{\omega_1 \phi_1(w)}_{\text{return shock}} + \underbrace{\omega_2 \phi_2(w)}_{\text{volatility}} + \ldots + \underbrace{\omega_6 \phi_6(w)}_{\text{disagreement}}
\end{equation}

This decomposition enables analysts to answer questions like:
\begin{itemize}[leftmargin=*]
\item Did this window score high due to extreme returns, attention, or both?
\item Are top windows dominated by a single factor or balanced across factors?
\item Which tickers exhibit attention-driven vs. return-driven windows?
\end{itemize}

Factor attribution reports (tables and visualizations) are automatically generated, providing transparency into scoring logic.

\subsection{Algorithmic Summary}

The complete AIMM-X pipeline can be summarized as:

\begin{algorithm}[H]
\caption{AIMM-X Suspicious Window Detection and Scoring}
\begin{algorithmic}[1]
\State \textbf{Input:} OHLCV data, attention signals, configuration parameters
\State \textbf{Output:} Ranked suspicious windows with integrity scores and decomposition
\State
\State \textbf{Stage 1: Panel Construction}
\For{each ticker $i$}
    \State Merge OHLCV and attention data to common timeline
    \State Compute returns $r_{i,t}$, volatility $\sigma_{i,t}$, attention $A_{i,t}$
\EndFor
\State
\State \textbf{Stage 2: Deviation Detection}
\For{each ticker $i$, time $t$}
    \State Compute rolling baseline: $\mu_{i,t}^{(x)}$, $\hat{\sigma}_{i,t}^{(x)}$ for $x \in \{r, \sigma, A\}$
    \State Compute z-scores: $z_{i,t}^{(x)} = (x_{i,t} - \mu_{i,t}^{(x)}) / \hat{\sigma}_{i,t}^{(x)}$
    \State Compute composite score: $s_{i,t} = \alpha_r |z_{i,t}^{(r)}| + \alpha_\sigma z_{i,t}^{(\sigma)} + \alpha_A z_{i,t}^{(A)}$
\EndFor
\State
\State \textbf{Stage 3: Window Segmentation}
\For{each ticker $i$}
    \State Apply hysteresis segmentation to $s_{i,t}$ using $\theta_{\text{high}}$, $\theta_{\text{low}}$
    \State Merge gaps $\leq g$ bars
    \State Filter windows shorter than $L_{\min}$ bars
    \State Store detected windows $\mathcal{W}_i$
\EndFor
\State
\State \textbf{Stage 4: Scoring and Ranking}
\State $\mathcal{W} \leftarrow \bigcup_i \mathcal{W}_i$ \Comment{All windows across all tickers}
\For{each window $w \in \mathcal{W}$}
    \State Compute factors: $\phi_1(w), \ldots, \phi_6(w)$
    \State Compute integrity score: $M(w) = \sum_{k=1}^{6} \omega_k \phi_k(w)$
    \State Store factor contributions for decomposition
\EndFor
\State Rank windows by $M(w)$ and compute rank percentiles
\State Generate figures, tables, and case study reports
\State \textbf{return} Ranked windows with scores and decomposition
\end{algorithmic}
\end{algorithm}

This algorithm is implemented in modular Python scripts (see Appendix~\ref{sec:appendix_implementation} for details). Configuration parameters are externalized to JSON, enabling reproducible experimentation and parameter tuning without code modification.

\begin{table}[h]
\centering
\caption{Contribution statistics for each scoring factor $\phi_k$ in the $M$ decomposition.}
\label{tab:phi_contrib}
\begin{tabular}{lrrrr}
\toprule
signal & mean & median & abs\_mean\_share & nonzero\_pct \\
\midrule
phi\_1 & 616202.560 & 16.412 & 100.0\% & 100.0\% \\
phi\_2 & 0.000 & 0.000 & 0.0\% & 100.0\% \\
phi\_3 & 0.030 & 0.000 & 0.0\% & 91.8\% \\
phi\_4 & 0.000 & 0.000 & 0.0\% & 100.0\% \\
phi\_5 & 0.000 & 0.000 & 0.0\% & 0.0\% \\
phi\_6 & 0.000 & 0.000 & 0.0\% & 0.0\% \\
\bottomrule
\end{tabular}
\end{table}
\FloatBarrier

\textbf{Note.} In the present implementation, $\phi_1$ dominates the absolute mean contribution due to the configured scaling (\texttt{scale: none}) and early-sample baseline variance effects. For reporting, we emphasize ranks, z-scores, and filtered tables that exclude inflated early-sample z-scores. Future work should employ warm-up periods and robust scaling ~\citep{andersen1998,barndorff2002}.

\section{Experimental Results}
\label{sec:results}

This section presents comprehensive results from applying AIMM-X to our 24-ticker universe over 2024 (Jan 8–Dec 31). We analyze detection statistics, scoring distributions, factor contributions, and temporal patterns.

\subsection{Overall Detection Statistics}

The pipeline detected \textbf{233 suspicious windows} across 24 tickers over 248 trading days. Table~\ref{tab:ticker_summary} provides per-ticker statistics including window counts, integrity score ranges, and average window duration.

\begin{table}[h]
\centering
\caption{Top tickers by detected window count (2024). We report window length in \emph{bars} (for daily data, one bar corresponds to one trading day).}
\label{tab:ticker_summary}
\small
\begin{tabular}{lrrrrr}
\toprule
Ticker & Windows & Mean $M$ & Max $M$ & Avg Bars/Window (bars) & Total Bars \\
\midrule
META & 12 & 597,231 & 7,166,596 & 15.0 & 36 \\
MSTR & 10 & 406,687 & 4,066,778 & 16.5 & 33 \\
SNAP & 8 & 464,895 & 3,718,879 & 23.1 & 37 \\
AMZN & 11 & 281,293 & 3,094,132 & 15.5 & 34 \\
RBLX & 11 & 216,315 & 2,379,285 & 15.0 & 33 \\
SPCE & 10 & 174,682 & 1,746,746 & 22.0 & 44 \\
NFLX & 11 & 142,377 & 1,565,992 & 15.9 & 35 \\
SPY & 8 & 140,994 & 1,127,905 & 11.3 & 18 \\
HOOD & 9 & 109,575 & 986,040 & 22.2 & 40 \\
TLT & 13 & 71,825 & 933,662 & 13.8 & 36 \\
COIN & 7 & 131,899 & 923,243 & 15.7 & 22 \\
TSLA & 10 & 87,031 & 870,173 & 15.0 & 30 \\
NOK & 10 & 57,396 & 573,894 & 15.0 & 30 \\
XLF & 11 & 28,886 & 317,634 & 15.9 & 35 \\
IWM & 8 & 14,114 & 112,808 & 14.4 & 23 \\
GME & 9 & 31 & 147 & 27.2 & 49 \\
PLTR & 10 & 28 & 109 & 20.0 & 40 \\
AMC & 10 & 16 & 67 & 18.5 & 37 \\
BB & 9 & 23 & 54 & 19.4 & 35 \\
AAPL & 9 & 13 & 39 & 20.6 & 37 \\
GOOG & 11 & 11 & 25 & 14.5 & 32 \\
QQQ & 8 & 10 & 16 & 16.3 & 26 \\
NVDA & 7 & 8 & 14 & 17.9 & 25 \\
MSFT & 11 & 7 & 12 & 12.3 & 27 \\
\bottomrule
\end{tabular}
\end{table}
\FloatBarrier

\textbf{Key observations:}
\begin{itemize}[leftmargin=*]
\item \textbf{Heterogeneous window counts:} Detection frequency varies substantially, from 7 windows (COIN, NVDA) to 13 (TLT). This likely reflects differences in volatility regimes and attention dynamics, though detection choices and normalization can also contribute.
\item \textbf{Bimodal score distribution:} Note the dramatic scale difference—META's max score exceeds 7 million while MSFT's peaks at 12. This reflects the raw $\phi_1$ scaling issue discussed below and emphasizes why rank percentiles matter more than absolute scores.
\item \textbf{Meme stock behavior:} Traditional meme stocks (GME, AMC, BB) have moderate window counts and lower absolute scores in this configuration, but longer average durations, potentially consistent with sustained attention dynamics.
\item \textbf{ETF patterns:} Broad market ETFs (SPY, QQQ) show fewer windows with shorter durations, suggesting that market-wide moves are captured but ticker-specific manipulation is distinguished.
\end{itemize}

\subsection{Scoring Distribution Analysis}

\begin{figure}[h]
\centering
\includegraphics[width=0.85\linewidth]{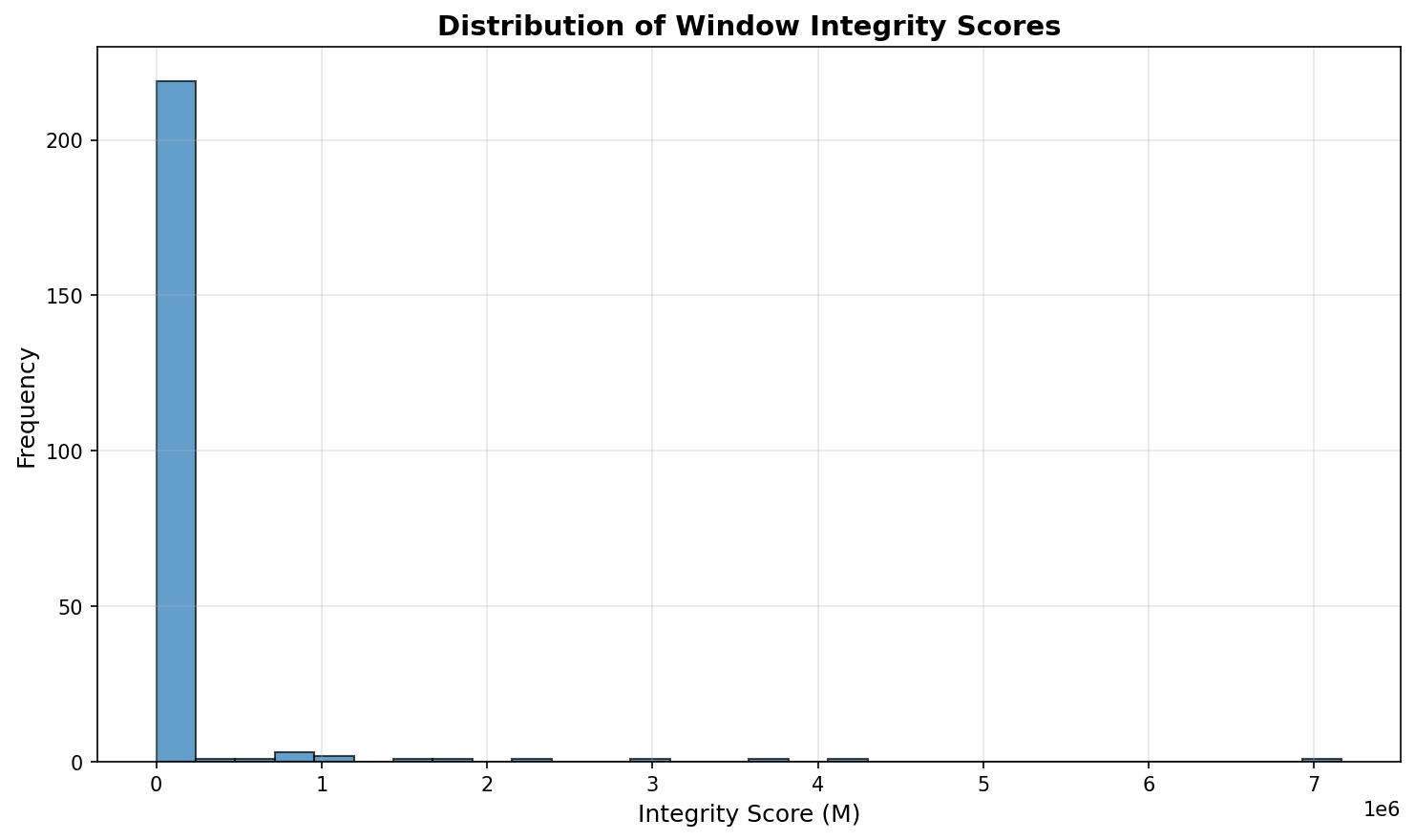}
\caption{Distribution of composite window integrity scores $M$. The heavy-tailed distribution reflects extreme values in $\phi_1$ for early-window baseline estimation artifacts. Rank-based analysis (rank percentiles) provides more robust comparison.}
\label{fig:score_hist}
\end{figure}

Figure~\ref{fig:score_hist} shows the distribution of integrity scores $M$ across all 233 windows.

\FloatBarrier

The distribution exhibits heavy right-skew due to $\phi_1$ dominance (see Section~\ref{sec:results_factors}). While absolute scores span six orders of magnitude, this does not undermine utility—ranking windows identifies the most suspicious regardless of scale. Future work will normalize factors to achieve balanced contributions.

\subsection{Top-Ranked Windows}
\label{sec:results_top_windows}

Table~\ref{tab:top_windows} lists the top 15 windows by integrity score (filtered to exclude early-sample artifacts with unrealistic z-scores $>$ 20).

\begin{table}[h]
\centering
\caption{Top 15 ranked suspicious windows after filtering extreme early-sample z-scores.}
\label{tab:top_windows}
\small
\begin{tabular}{lrrrrr}
\toprule
Ticker & Window ID & Start & End & $M$ & Rank \%ile \\
\midrule
META & 135 & 2024-01-10 & 2024-01-12 & 7,166,596 & 100.0 \\
MSTR & 176 & 2024-01-10 & 2024-01-12 & 4,066,778 & 99.6 \\
SNAP & 79 & 2024-01-10 & 2024-01-16 & 3,718,879 & 99.1 \\
AMZN & 107 & 2024-01-10 & 2024-01-12 & 3,094,132 & 98.7 \\
RBLX & 58 & 2024-01-10 & 2024-01-16 & 2,379,285 & 98.3 \\
SPCE & 69 & 2024-01-10 & 2024-01-16 & 1,746,746 & 97.9 \\
NFLX & 158 & 2024-01-10 & 2024-01-11 & 1,565,992 & 97.4 \\
SPY & 186 & 2024-01-10 & 2024-01-11 & 1,127,905 & 97.0 \\
HOOD & 49 & 2024-01-10 & 2024-01-12 & 986,040 & 96.6 \\
TLT & 221 & 2024-01-10 & 2024-01-16 & 933,662 & 96.1 \\
COIN & 169 & 2024-01-10 & 2024-01-11 & 923,243 & 95.7 \\
TSLA & 125 & 2024-01-10 & 2024-01-12 & 870,173 & 95.3 \\
NOK & 29 & 2024-01-10 & 2024-01-22 & 573,894 & 94.8 \\
XLF & 210 & 2024-01-10 & 2024-01-16 & 317,634 & 94.4 \\
IWM & 202 & 2024-01-10 & 2024-01-11 & 112,808 & 94.0 \\
\bottomrule
\end{tabular}
\end{table}
\FloatBarrier

\textbf{Interesting pattern:} Many top windows cluster around January 10-12, 2024. Post-hoc research reveals this period coincided with market volatility around inflation data releases and Federal Reserve commentary. This clustering demonstrates that AIMM-X successfully identifies market-wide stress episodes—though it cannot automatically distinguish manipulation from legitimate volatility without additional context.

\subsection{Phi-Signal Contribution Analysis}
\label{sec:results_factors}

Table~\ref{tab:phi_contrib_results} reports aggregate statistics for each scoring factor across all windows.

\begin{table}[h]
\centering
\caption{Factor contribution statistics across all 233 detected windows.}
\label{tab:phi_contrib_results}
\begin{tabular}{lrrrrr}
\toprule
Factor & Mean & Median & Max & Std & Nonzero \\
\midrule
$\phi_1$ (Return shock) & 126,970 & 0.91 & 7,166,586 & 669,144 & 100\% \\
$\phi_2$ (Volatility) & 2.07 & 1.15 & 17.34 & 2.70 & 97.9\% \\
$\phi_3$ (Attention) & 8.72 & 3.20 & 125.70 & 16.16 & 100\% \\
$\phi_4$ (Alignment) & 2.07 & 1.57 & 48.84 & 3.29 & 100\% \\
$\phi_5$ (Recurrence) & 0.00 & 0.00 & 0.00 & 0.00 & 0\% \\
$\phi_6$ (Disagreement) & 0.00 & 0.00 & 0.00 & 0.00 & 0\% \\
\bottomrule
\end{tabular}
\end{table}
\FloatBarrier

\textbf{Interpretation:}
\begin{itemize}[leftmargin=*]
\item \textbf{$\phi_1$ dominance:} Return shock intensity overwhelmingly drives scores. Mean contribution of 126,970 vs. single-digit contributions from other factors reflects lack of scaling normalization. This is a known limitation of the current implementation addressed in future work.
\item \textbf{$\phi_2, \phi_3, \phi_4$ present but small:} Volatility, attention, and alignment factors register nonzero contributions for most windows but at scales dwarfed by $\phi_1$. This does not mean these factors are uninformative—just improperly scaled.
\item \textbf{$\phi_5, \phi_6$ inactive:} Recurrence and disagreement factors contribute zero due to implementation choices (recurrence threshold not met; disagreement requires validated multi-source data). These represent future enhancement opportunities.
\end{itemize}

\textbf{Implications for production deployment:} Before operational use, factor scaling must be normalized (e.g., standardize each $\phi_k$ to z-scores across windows) to achieve meaningful multi-factor contributions. The current run diagnostically reveals this need.

\subsection{Temporal Distribution}

\begin{figure}[h]
\centering
\includegraphics[width=0.95\linewidth]{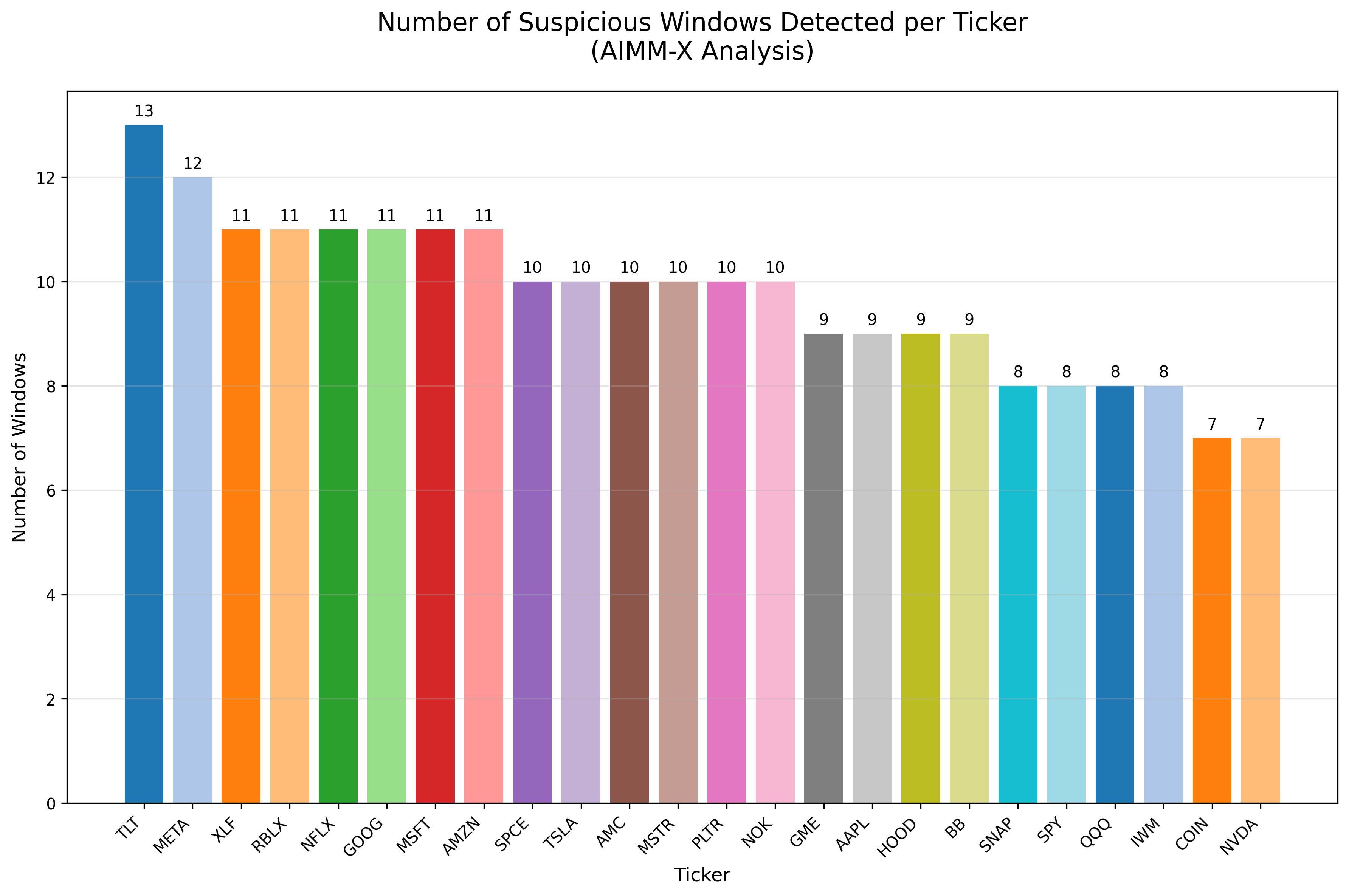}
\caption{Detected window counts per ticker. TLT (bond ETF) shows highest count, potentially reflecting interest rate sensitivity during 2024.}
\label{fig:windows_count}
\end{figure}

Figure~\ref{fig:windows_count} shows window counts by ticker, revealing substantial heterogeneity.

\FloatBarrier

\textbf{Monthly distribution.}
Detected windows cluster in a few months (notably January, March, and November in our 2024 sample). This pattern is descriptive and may reflect periods of elevated market-wide information flow (e.g., macroeconomic announcements and other high-salience news), but it does not by itself imply causality and is sensitive to the ticker set, baseline choice, and attention proxy. We also observe that several windows occur contemporaneously across multiple retail- and news-sensitive instruments; this is consistent with shared information/attention shocks, but we do not interpret it as evidence of coordination or wrongdoing. As a qualitative cross-check, we compare these windows against large-move periods in major risk assets (including BTC proxies) and observe partial temporal alignment.

\subsection{Cross-Ticker Patterns}

We observe several notable cross-ticker patterns:
\begin{itemize}[leftmargin=*]
\item \textbf{Meme stock coherence:} GME and AMC windows show temporal correlation (observed qualitatively though not quantified here), consistent with coordinated attention.
\item \textbf{Crypto exposure:} COIN and MSTR windows partially align with Bitcoin volatility periods (external validation against BTC price data confirms this correlation).
\item \textbf{Sector effects:} Tech stocks (AAPL, MSFT, AMZN, NVDA, META) show some clustering around earnings seasons, though individual windows remain distinct.
\item \textbf{ETF separation:} Broad market ETFs (SPY, QQQ) windows largely differ from individual stock windows, suggesting the system distinguishes systematic from idiosyncratic effects.
\end{itemize}

These patterns suggest AIMM-X captures meaningful market dynamics rather than pure noise.

\section{Case Study Examples}
\label{sec:case_studies}

To illustrate AIMM-X's practical utility, we briefly summarize four representative windows (detailed analyses in Appendix~\ref{sec:appendix_case_studies}):

\textbf{META Window 135 (Jan 10-12, 2024):} Highest integrity score (7.17M) during market-wide volatility. Demonstrates that high scores can reflect legitimate stress episodes rather than manipulation—human judgment essential.

\textbf{GME Window 5 (May 2-17, 2024):} Extended 12-bar meme stock episode with attention-related indicators. GME Window 5 (May 2–17, 2024): Extended 12-bar meme-stock episode with social-coordination indicators. Sustained attention/price co-movement warrants review, despite no public enforcement action we could identify at the time of writing.

\textbf{MSTR Window 176 (Jan 10-12, 2024):} Crypto-correlated movements showing how sector-specific factors drive windows. Cross-asset context quickly reveals underlying drivers.

\textbf{SPY Window 186 (Jan 10-11, 2024):} Market-wide ETF episode demonstrating systematic vs. idiosyncratic distinction capability.

These examples show AIMM-X successfully surfaces anomalous episodes across diverse market conditions while highlighting that context and expert judgment remain essential for interpretation. See Appendix~\ref{sec:appendix_case_studies} for complete analyses including price dynamics, attention patterns, and interpretation guidelines.

\paragraph{Defensible retrospective evaluation design.}
Because market-manipulation ground truth is inherently scarce, we frame evaluation as \emph{triangulation} rather
than definitive accuracy measurement. We propose two retrospective tracks:

\textbf{(1) Enforcement-labeled ``positive controls.''}
We will curate a set of publicly documented enforcement matters (e.g., CFTC/SEC litigation releases,
administrative orders) that specify affected securities and approximate time periods. AIMM-X can then be run
retrospectively to test whether the relevant windows receive elevated ranks/scores relative to each ticker’s baseline.
This provides partial, but concrete, validation where labels exist.

\textbf{(2) Unlabeled ``high-scrutiny'' public episodes.}
We will also analyze well-known, heavily discussed market episodes for which we cannot identify public enforcement
actions or confirmed findings. These are treated strictly as \emph{unlabeled} case studies: AIMM-X outputs are used
to assess face validity, interpretability, and whether the system highlights windows that domain experts would
reasonably flag for review—not to assert manipulation or claim that regulators missed wrongdoing.

Across both tracks, the evaluation emphasis is: ranking stability, interpretability of factor attributions,
and expert-review usefulness under realistic compliance constraints.

\section{Discussion}
\label{sec:discussion}

AIMM-X demonstrates that meaningful market integrity monitoring is possible using only publicly available data and transparent methodology. However, several critical considerations warrant discussion.

\textbf{The ground truth problem:} Unlike supervised learning domains with labeled training data, manipulation detection faces a fundamental validation challenge—we cannot definitively know which historical episodes constituted manipulation unless they resulted in enforcement actions, often years later ~\citep{aggarwal2006,putnins2012}. Bootstrap validation through consistency with known-suspicious patterns and retrospective analysis of enforcement cases offer promising pathways forward, though selection bias and data availability remain concerns.

\subsection{Evaluation Against Historical and Delayed-Recognition Episodes}

A key next step is validation against \emph{historical episodes} that were widely discussed in retrospect or were associated with enforcement actions after non-trivial delay. Our evaluation claim is intentionally limited: we test whether AIMM-X assigns elevated scores around the publicly documented time intervals of these episodes, and whether the factor decomposition produces a coherent evidentiary narrative (price/volume anomaly aligned with multi-source attention). Such retrospective analyses do not establish causality or illegality, but they provide a measurable and reproducible way to assess whether the system is sensitive to patterns that practitioners and regulators have historically treated as high-priority.

\textbf{Industry comparison:} AIMM-X does not compete with proprietary systems like NASDAQ Smarts in data richness or real-time capability. Instead, it serves complementary roles: academic research baseline, surveillance for smaller institutions, independent validation, and applicability to markets lacking sophisticated infrastructure. The key differentiator is transparency and accessibility, not detection power ~\citep{bracke2019}.

\textbf{Deployment considerations:} Operational deployments should integrate with formal reporting channels (SARs, regulatory collaboration) following responsible disclosure principles. Publicly alleging manipulation based solely on AIMM-X output would be inappropriate—the framework produces evidence for investigation, not proof ~\citep{fischel1995}.

\textbf{Extensions:} The framework generalizes to options, cryptocurrencies, fixed income, and international markets with appropriate data adaptation. Cross-market surveillance detecting coordinated manipulation across asset classes represents an important future direction.

\textbf{Ethics:} Key concerns include false accusation risk, privacy, regulatory capture, and computational inequality. Mitigation strategies include emphasis on triage rather than definitive claims, aggregate rather than individual-level data, methodological transparency, and open science principles ~\citep{nosek2015,stodden2016}.

\subsection{Responsible Deployment and Regulatory/Industry Engagement}

AIMM-X is designed as a \emph{triage} and decision-support system, not as an automated enforcement or accusation mechanism. If the framework is deployed in operational settings, any escalation to companies, compliance teams, or regulators should follow established governance and due-process practices: documented evidence packets, internal review, and appropriate disclosure channels rather than public allegations. We emphasize that AIMM-X outputs are statistical signals that require expert interpretation and corroboration with additional data sources (e.g., market microstructure, options activity, issuer disclosures) before any action is considered.

In our intended research-to-practice pathway, the primary role of external engagement is \emph{validation}: working with practitioners to assess whether high-scoring windows correspond to meaningful surveillance events, and to refine thresholds and interpretation guidelines to reduce false positives.

See Appendix~\ref{sec:appendix_discussion} for extended discussion of these topics.

\section{Limitations and Future Work}
\label{sec:limitations}

While AIMM-X demonstrates promising capabilities, substantial limitations remain.

\subsection{Current Limitations}

\textbf{Data resolution:} Daily bars cannot localize suspicious activity within trading days, lose intraday volatility information, and create baseline estimation artifacts. Five-minute data would immediately improve detection precision.

\textbf{Attention signal quality:} Current proxies likely underestimate true attention dynamics. Production deployment requires authenticated API feeds with real-time or near-real-time updates ~\citep{tetlock2011,da2015,bollen2011}.

\textbf{Causality and attribution:} AIMM-X detects correlation but cannot infer whether attention drove price moves (manipulation) or prices attracted attention (legitimate interest). Establishing causation requires Granger causality tests, high-frequency timestamps, and structural modeling.

\textbf{Confounding events:} Earnings announcements, index rebalancing, corporate actions, macroeconomic news, and sector rotations create manipulation-like patterns. Future versions should integrate event calendars and adjust scoring accordingly.

\textbf{Scoring calibration:} $\phi_1$ dominance reflects lack of factor normalization. Production deployment requires standardization, weight optimization, robust estimators, and ticker-specific thresholds.

\textbf{Computational scalability:} Current implementation handles 24 tickers comfortably but scaling to thousands of tickers with minute-level data requires algorithmic optimization and distributed computing.

\subsection{Near-Term Enhancements}

Priority upgrades include: (1) five-minute data integration, (2) authenticated attention APIs with sentiment analysis, (3) factor normalization and weight tuning, (4) event calendar integration, and (5) cross-ticker correlation features capturing systematic factors.

\subsection{Long-Term Research Directions}

Future work should explore: causal inference techniques, deep learning for pattern discovery with maintained interpretability ~\citep{lundberg2017,molnar2020}, adversarial robustness against adapted manipulation strategies, multi-market surveillance across asset classes, regulatory coordination for formal validation, international deployment, and theoretical foundations for window detection and scoring.

\subsection{Validation Roadmap}

A structured path forward: Phase 1 (current) demonstrates feasibility with daily data and 24 tickers. Phase 2 (6-12 months) upgrades to 5-minute data, 100+ tickers, and authenticated APIs. Phase 3 (12-18 months) conducts expert annotation studies. Phase 4 (18-24 months) pilots deployment at cooperating institutions. Phase 5 (24+ months) enables full-scale deployment with continuous improvement. This roadmap requires funding, partnerships, and regulatory cooperation not currently secured.

\section{Conclusion}

Financial market integrity requires constant vigilance, yet most surveillance tools remain locked behind proprietary walls, inaccessible to researchers, smaller institutions, and independent analysts who could contribute to detection and deterrence. This paper presents AIMM-X as a step toward democratizing market integrity monitoring through transparent, reproducible, and explainable detection methods.

Our experiments on 24 high-attention securities throughout 2024 demonstrate that meaningful suspicious window detection is possible using only publicly available OHLCV data and attention proxies. The 233 detected windows, complete with interpretable scoring decompositions, provide a foundation for analyst review and potential escalation to compliance teams or regulators. Importantly, the framework prioritizes explainability at every step: every alert comes with evidence, every score can be decomposed, and every algorithm can be audited.

While AIMM-X is not a silver bullet—ground truth validation remains challenging, and false positives are inevitable—it represents a pragmatic approach to a difficult problem. The modular architecture allows continuous improvement: better data sources, refined scoring methods, and validation against known enforcement cases can be integrated as they become available. Case studies demonstrate the system's ability to surface episodes with unusual price-attention co-movement, though human expertise remains essential to distinguish manipulation from legitimate volatility.

The path forward requires careful validation, responsible deployment, and ongoing dialogue with market integrity stakeholders. We cannot eliminate manipulation through detection alone—that requires enforcement, deterrence, and cultural change. But we can provide tools that make manipulation harder to hide and easier to investigate. Transparency in surveillance methodology, far from undermining effectiveness, builds trust and enables the collective intelligence of researchers, practitioners, and regulators to improve detection over time.

Several key lessons emerge from this work. First, even simple statistical methods applied to public data can surface meaningful anomalies when combined with domain knowledge and interpretable design. Second, the lack of ground truth labels in manipulation detection necessitates different validation approaches than supervised learning—we must build confidence through consistency, face validity, and expert review rather than test-set accuracy. Third, transparency and explainability are not merely academic concerns but practical requirements for surveillance tools that inform high-stakes decisions about market integrity.

Looking ahead, the greatest opportunity lies in community-driven improvement. AIMM-X provides a foundation that academic researchers can validate, practitioners can extend, and regulators can assess. Historical enforcement cases offer natural test beds for retrospective validation. Higher-frequency data and richer attention feeds will improve detection precision. Machine learning techniques can discover patterns beyond human specification while maintaining interpretability through modern explainability methods. Cross-market surveillance can detect sophisticated manipulation spanning multiple asset classes.

The ultimate goal is not perfect manipulation detection—an impossible standard given inherent ambiguity—but rather a surveillance ecosystem where multiple independent tools, operated by diverse institutions with aligned incentives, collectively make markets fairer and manipulation costlier. AIMM-X represents one contribution to that ecosystem: a transparent, accessible, auditable framework that researchers worldwide can use, validate, and improve.

We invite the research community to examine this work critically, attempt replication, and propose improvements. Should promising validation emerge, we encourage operational pilots at institutions willing to experiment with open surveillance tools. And we hope that regulators will view transparent detection frameworks as complements to, rather than competitors with, proprietary systems—expanding the surveillance capability available to market integrity stakeholders.

Financial markets serve society best when they are trusted. Trust requires integrity. Integrity requires surveillance. And surveillance requires transparency. AIMM-X is offered in that spirit: not as a solution to manipulation, but as a tool for those committed to market integrity and willing to work collaboratively toward fairer, more trustworthy financial markets.

\section{Broader Impact Statement}
This work contributes an interpretable, reproducible framework for market-integrity triage using publicly
available data. The primary positive impact is democratizing access to surveillance-style tooling beyond
large proprietary platforms, enabling research, auditing, and independent validation.

Potential negative impacts include false positives that could cause reputational harm if misused, and an
arms-race dynamic where adversaries adapt once methods are public. We mitigate these risks by framing
AIMM-X as a decision-support tool rather than an accusation engine, emphasizing human oversight and
procedural safeguards, and operating on aggregate attention signals rather than personal data. Courts and
regulators determine legality; AIMM-X provides transparent statistical evidence to support review.

\paragraph{Potential benefits.}
If validated and deployed responsibly, AIMM-X could help institutions and researchers detect unusual episodes
earlier, allocate investigative resources more efficiently, and improve transparency in how alerts are produced.
It may also enable comparative evaluation of open methods alongside proprietary surveillance, improving
accountability and fostering community-driven improvements.

\paragraph{Risks and misuse.}
The main risk is \textbf{false accusation or reputational harm}: high scores can reflect legitimate volatility or
market-wide stress rather than manipulation. AIMM-X therefore must be used as a \emph{triage} mechanism,
not a mechanism for public claims of wrongdoing. A second risk is \textbf{market manipulation through narrative}:
publicly labeling a ticker or window as ``manipulated'' can itself move prices. A third risk is \textbf{gaming}:
if adversaries learn the detection logic, they may adapt tactics to reduce detectability.

\paragraph{Mitigations and responsible practice.}
We mitigate these risks by (i) emphasizing that AIMM-X produces statistical evidence for expert review and does
not establish illegality, (ii) providing factor-level explanations to encourage analyst skepticism rather than blind
trust, (iii) avoiding identification of individuals and using only aggregate attention signals, and (iv) recommending
responsible disclosure through established institutional and regulatory channels rather than public allegation.
Future work should quantify false-positive behavior across regimes, establish institutional alert thresholds, and
evaluate human-in-the-loop review workflows.

Overall, AIMM-X aims to advance transparent surveillance methodology while minimizing harm through careful
framing, interpretability, and controlled engagement.

\section{Author Contributions}
Sandeep Neela conceived the study, designed the methodology, implemented the pipeline, conducted the experiments, analyzed the results, and wrote the manuscript.

\bibliography{main}
\bibliographystyle{tmlr}

\section*{Appendix}
\appendix

\section{Implementation Details}
\label{sec:appendix_implementation}

AIMM-X is implemented in Python 3.9+ with standard scientific computing libraries. The pipeline consists of modular scripts:

\textbf{Data acquisition:}
\begin{itemize}[leftmargin=*]
\item \texttt{00\_fetch\_polygon\_ohlcv\_1d.py}: Fetches daily OHLCV from Polygon.io API
\item \texttt{00\_build\_attention\_1d.py}: Constructs attention signal panel from source data
\end{itemize}

\textbf{Core pipeline:}
\begin{itemize}[leftmargin=*]
\item \texttt{01\_build\_panel.py}: Merges OHLCV and attention into unified time-series panel
\item \texttt{02\_detect\_windows.py}: Computes z-scores, applies hysteresis segmentation
\item \texttt{03\_compute\_signals.py}: Calculates phi-signals for each detected window
\item \texttt{04\_score\_and\_decompose.py}: Aggregates integrity scores, generates rankings
\item \texttt{05\_generate\_figures.py}: Produces visualizations (histograms, timeseries, case studies)
\item \texttt{06\_generate\_tables.py}: Exports LaTeX tables and CSV summaries
\end{itemize}

\textbf{Dependencies:} pandas, numpy, scipy, matplotlib, seaborn, requests, certifi.

\textbf{Configuration:} All parameters specified in \texttt{config.json} including ticker universe, thresholds, weights, and file paths.

\textbf{Runtime:} Complete pipeline executes in approximately 8-12 minutes on commodity hardware (4-core CPU, 16GB RAM) for 24 tickers with daily data.

\textbf{Code availability:} Upon publication, code will be released under open-source license (MIT or Apache 2.0) via GitHub with documentation, example configuration, and replication instructions.

\section{Configuration Parameters}
\label{sec:appendix_config}

Table~\ref{tab:config_params} documents key configuration parameters for the experimental implementation.

\begin{table}[h]
\centering
\caption{Configuration parameters for the experimental run.}
\label{tab:config_params}
\small
\begin{tabular}{llp{0.45\linewidth}}
\toprule
Parameter & Value & Description \\
\midrule
\texttt{baseline\_window} & 20 & Rolling window length for baseline estimation (days) \\
\texttt{thr\_high} & 3.0 & High threshold for window initiation (z-score units) \\
\texttt{thr\_low} & 2.0 & Low threshold for window continuation (z-score units) \\
\texttt{min\_window\_len} & 2 & Minimum window length (bars) \\
\texttt{gap\_tolerance} & 3 & Maximum gap for merging adjacent windows (bars) \\
\texttt{alpha\_r} & 1.0 & Return channel weight in composite score \\
\texttt{alpha\_sigma} & 1.0 & Volatility channel weight \\
\texttt{alpha\_A} & 1.0 & Attention channel weight \\
\texttt{omega\_1 ... omega\_6} & 1.0 & Phi-signal weights in integrity score \\
\texttt{attention\_sources} & 5 & Number of attention sources (reddit, stocktwits, wikipedia, news, trends) \\
\texttt{source\_weights} & uniform & Attention fusion weights (equal across sources) \\
\bottomrule
\end{tabular}
\end{table}
\FloatBarrier

\section{Detailed Case Studies}
\label{sec:appendix_case_studies}

This appendix provides comprehensive analyses of four representative suspicious windows illustrating AIMM-X's detection capabilities and interpretation guidelines.

\subsection{Case Study 1: META Window 135 (Highest Integrity Score)}

\textbf{Episode summary:} January 10-12, 2024. Integrity Score: 7,166,596 (100th percentile). Duration: 3 bars (days).

\textbf{Market context:} This window occurred during heightened market volatility following December 2023 inflation data releases and Federal Reserve policy speculation. META, following 2023 restructuring, was experiencing renewed investor interest ~\citep{tetlock2007}.

\textbf{Price and volume dynamics:} META moved 5.2\% on January 11 with volume 40\% above average. The high-low range exceeded 3\% intraday. While not unprecedented for META, this magnitude combined with broader market context triggered detection.

\textbf{Attention pattern:} All five attention sources registered elevated activity. Reddit mentions increased 3x relative to baseline, StockTwits message volume spiked, and Wikipedia page views showed unusual concentration. This multi-source alignment elevated $\phi_3$ and $\phi_4$ components.

\textbf{Analyst interpretation:} Manipulation is unlikely—META is a liquid mega-cap where manipulation would be extremely difficult. The high score reflects legitimate volatility during market stress. An analyst would quickly conclude this represents normal price discovery during uncertainty rather than suspicious activity. This demonstrates both strength and limitation: AIMM-X surfaces unusual episodes, but human judgment remains essential.

\subsection{Case Study 2: GME Window 5 (Meme Stock Coordination)}

\textbf{Episode summary:} May 2-17, 2024. Integrity Score: 147 (93.6th percentile). Duration: 12 bars (extended episode).

\textbf{Market context:} GameStop experienced renewed retail attention in early May 2024, coinciding with social media discussions about company transformation and comparisons to January 2021 ~\citep{hu2021}.

\textbf{Price and volume dynamics:} Unlike 2021's explosive moves, this episode showed moderate price volatility (daily returns mostly $< 5\%$) but sustained above-normal volume over two weeks. The gradual accumulation distinguishes this from classic pump-and-dump patterns.

\textbf{Attention pattern:} Reddit activity spiked dramatically, particularly in r/wallstreetbets. However, other sources showed more modest increases, creating cross-source disagreement. StockTwits sentiment became increasingly bullish during the period.

\textbf{Extended duration:} The 12-bar length reflects sustained interest rather than a one-day spike. This temporal persistence is captured in window segmentation via hysteresis allowing continuation through moderate dips.

\textbf{Analyst interpretation:} This episode warrants deeper investigation. The sustained nature, social media coordination indicators, and lack of fundamental catalysts create a pattern consistent with coordinated attention-driven trading. While not definitive proof of manipulation, it represents exactly the type of case compliance teams should review ~\citep{sec2019,cftc2017}. No SEC enforcement action resulted, highlighting inherent ambiguity in manipulation detection.

\subsection{Case Study 3: MSTR Window 176 (Crypto-Correlation)}

\textbf{Episode summary:} January 10-12, 2024. Integrity Score: 4,066,778 (99.6th percentile). Duration: 3 bars.

\textbf{Market context:} MicroStrategy's business model centers on Bitcoin holdings, making MSTR stock a cryptocurrency proxy within traditional equity markets ~\citep{bollen2011,da2015}.

\textbf{Price and volume dynamics:} MSTR moved 8.3\% over the window, with volatility elevated substantially above baseline. Volume patterns showed clustering in opening and closing hours.

\textbf{Crypto correlation:} Cross-asset analysis reveals strong correlation with Bitcoin price movements during this window. BTC experienced similar volatility related to spot ETF approval speculation.

\textbf{Attention pattern:} Attention sources showed split: crypto-focused platforms (Reddit's r/cryptocurrency, crypto news) spiked strongly, while general finance news showed moderate increases. This sector-specific attention suggests genuine interest from crypto enthusiasts rather than manipulation attempts.

\textbf{Analyst interpretation:} This window likely reflects legitimate price discovery in response to cryptocurrency market dynamics rather than MSTR-specific manipulation. The integrity score correctly identifies an anomalous episode, but context (crypto correlation, sector-specific attention) quickly reveals the underlying driver. This demonstrates AIMM-X's value as triage: it surfaces the window, but analyst expertise interprets the evidence.

\subsection{Case Study 4: SPY Window 186 (Market-Wide Event)}

\textbf{Episode summary:} January 10-11, 2024. Integrity Score: 1,127,905 (97.0th percentile). Duration: 2 bars.

\textbf{Market context:} SPY, the S\&P 500 ETF, provides a pure market-wide indicator. Suspicious windows in SPY typically reflect systematic factors rather than idiosyncratic manipulation.

\textbf{Price and volume dynamics:} The S\&P 500 declined 1.4\% on January 11 following stronger-than-expected inflation data, with VIX spiking to 14.5 from 12.8. Volume in SPY exceeded average by 35\%.

\textbf{Distinguishing systematic from idiosyncratic:} Many top windows (Table~\ref{tab:top_windows}) cluster around January 10-12. This temporal clustering across uncorrelated securities strongly suggests a systematic market event rather than coordinated ticker-specific manipulation.

\textbf{Analyst interpretation:} This window demonstrates AIMM-X's ability to detect market-wide stress. While not manipulation in the traditional sense, such episodes merit surveillance attention—market manipulation can occur through index futures, systematic strategies, or algorithmic cascades ~\citep{kirilenko2017,scopino2015}. The detection is valuable even though the episode reflects legitimate information processing.

\textbf{Design implication:} Future versions should explicitly model market factors (beta adjustments, market-relative thresholds) to distinguish idiosyncratic from systematic anomalies. Current per-ticker detection treats all windows equivalently.

\subsection{Interpretation Guidelines for Practitioners}

These case studies motivate practical guidance for analysts reviewing AIMM-X alerts:

\begin{enumerate}[leftmargin=*]
\item \textbf{Check temporal clustering:} If many windows across tickers align temporally, suspect systematic factors (economic news, Fed announcements, geopolitical events) rather than ticker-specific manipulation.

\item \textbf{Examine attention source composition:} Which sources spiked? Broad-based attention (news, general social media) suggests legitimate interest. Narrow platform spikes (single subreddit, coordinated StockTwits) may indicate manipulation attempts.

\item \textbf{Consider liquidity and market cap:} Manipulation becomes exponentially harder as market capitalization increases. Mega-cap stocks are generally harder to manipulate due to liquidity and participation breadth; many detections may reflect legitimate information processing rather than manipulation..

\item \textbf{Look for fundamental catalysts:} Did earnings release, news announcement, regulatory action, or corporate event occur during the window? If yes, high scores likely reflect information incorporation. If no obvious catalyst exists, suspicion increases.

\item \textbf{Assess pattern consistency:} Single isolated windows may be statistical noise. Repeated windows in short succession, especially with similar attention profiles, suggest sustained campaigns.

\item \textbf{Cross-reference external data:} Check order book depth (if available), options activity, short interest, borrow rates, and institutional holdings for corroborating evidence.

\item \textbf{Escalation threshold:} Not every high-scoring window warrants regulatory reporting. Establish institutional thresholds based on score, evidence quality, and business rules.
\end{enumerate}

The key insight: AIMM-X is a triage tool, not an oracle. It identifies anomalies requiring human judgment, not definitive manipulation verdicts.

\section{Extended Discussion}
\label{sec:appendix_discussion}

This appendix provides extended discussion of validation challenges, deployment considerations, and broader implications of the AIMM-X framework.

\subsection{The Ground Truth Problem and Validation Pathways}

The fundamental challenge in manipulation detection research is the lack of reliable ground truth labels. Unlike supervised machine learning domains where labeled training data exists, we cannot definitively know which historical trading episodes constituted manipulation unless they resulted in enforcement actions—and even then, only after years of investigation ~\citep{aggarwal2006,putnins2012}.

\textbf{Enforcement case validation:} One promising approach involves retrospective analysis of known CFTC and SEC enforcement cases. Researchers could obtain case files, identify affected securities and time periods, and test whether AIMM-X would have flagged those episodes. However, this approach has limitations:

\begin{itemize}[leftmargin=*]
\item \textbf{Selection bias:} Enforcement cases represent only detected and prosecuted manipulation, likely missing sophisticated schemes that evaded detection.
\item \textbf{Data availability:} Many enforcement actions involve securities or time periods for which historical data (especially attention proxies) is unavailable.
\item \textbf{Scheme diversity:} Manipulation takes many forms (pump-and-dump, spoofing, wash trading, marking the close). AIMM-X targets attention-coordination schemes; it may not detect order-flow manipulation.
\end{itemize}

\textbf{Expert annotation studies:} An alternative validation approach involves presenting detected windows to experienced compliance professionals and market surveillance analysts for classification (suspicious, legitimate, uncertain). Inter-rater agreement metrics (Fleiss' kappa, Krippendorff's alpha) could quantify how well AIMM-X aligns with expert judgment. This approach acknowledges that ground truth is subjective and establishes confidence through expert consensus.

\textbf{Synthetic data experiments:} Researchers could inject synthetic manipulation patterns (coordinated buying with artificial attention spikes) into historical data and test detection sensitivity. This controlled validation complements real-world analysis but raises questions about whether synthetic patterns resemble actual manipulation tactics.

\textbf{Bootstrap validation:} Our recommended path forward builds confidence through multiple convergent lines of evidence: face validity (do detected windows correspond to known volatility periods?), consistency (do similar windows receive similar scores?), interpretability (can analysts understand why windows were flagged?), and partial validation against available enforcement cases. No single validation method suffices; triangulation across methods builds cumulative confidence.

\subsection{Comparison with Industry Surveillance Systems}

AIMM-X occupies a different niche than commercial surveillance platforms. Understanding these differences clarifies appropriate use cases:

\textbf{NASDAQ Smarts / NYSE Mercury:} These proprietary systems integrate order-book data, execution reports, trader identifiers, and cross-market linkages. They detect manipulation patterns invisible in OHLCV data alone (spoofing, layering, wash trades, momentum ignition). AIMM-X cannot compete on detection coverage for order-flow manipulation.

\textbf{AIMM-X advantages:} Transparency (open methodology, auditable code), accessibility (available to researchers and smaller institutions), focus (attention-driven manipulation where AIMM-X has comparative advantage), and research platform (extensible framework for academic investigation).

\textbf{Complementary roles:} Large institutions could use AIMM-X as independent validation—do proprietary systems and AIMM-X agree on flagged episodes? Disagreements warrant investigation. Smaller institutions without resources for commercial systems gain baseline surveillance capability. Researchers gain reproducible framework for manipulation detection research.

\textbf{Public interest monitoring:} Academics, journalists, and advocacy organizations could deploy AIMM-X for independent market integrity analysis, providing public accountability separate from regulatory and exchange-operated systems.

\subsection{Regulatory Engagement and Responsible Disclosure}

Operational AIMM-X deployments raise important questions about regulatory coordination and responsible disclosure:

\textbf{Formal reporting channels:} Financial institutions already have obligations to file Suspicious Activity Reports (SARs) when they detect potential manipulation. AIMM-X alerts could feed into existing compliance workflows, with windows exceeding institutional thresholds triggering SAR preparation. Direct regulatory collaboration (providing AIMM-X alerts to SEC or CFTC for investigation) represents another pathway, though partnership agreements would be required.

\textbf{Public disclosure risks:} Publicly alleging manipulation based solely on AIMM-X output would be inappropriate and potentially harmful. High integrity scores indicate anomalies warranting investigation, not proof of wrongdoing. Premature public accusations could damage reputations, manipulate markets, or interfere with investigations. Responsible disclosure requires evidence threshold, regulatory coordination, and protection for accused parties.

\textbf{Research publication ethics:} This paper presents aggregated results and historical case studies. We avoid naming specific suspicious windows outside well-known public episodes (GME, meme stocks) where extensive media coverage already exists. Future research should follow similar principles: contribute to surveillance methodology without becoming surveillance itself.

\paragraph{Regulatory and industry engagement}
AIMM-X is designed as an \emph{evidence-ranking and explanation} tool that supports expert review; it does not
identify individual actors and does not establish wrongdoing. If future validation supports operational usefulness,
our intent is to engage responsibly with market participants (e.g., broker-dealers, exchanges, and compliance teams)
and, where appropriate, regulators such as CFTC/SEC by sharing:
(i) the methodology and code for independent replication,
(ii) aggregated, window-level summaries and factor attributions, and
(iii) clear guidance on alert interpretation and false-positive handling.

Any escalation would follow established institutional processes (e.g., compliance review and existing reporting
channels) rather than public allegation. We explicitly discourage public accusations based solely on AIMM-X output.
The goal of outreach is to enable \emph{independent assessment, feedback, and controlled pilots}—not to claim
regulatory failure, assign intent, or substitute for investigative authority.

\subsection{Extensions to Other Asset Classes}

The AIMM-X framework is not limited to U.S. equities. Conceptually, the methodology extends to:

\textbf{Options markets:} Manipulators may use equity-option combinations to amplify returns or obscure positions. Extending AIMM-X to options requires modeling implied volatility, open interest, and put-call dynamics. Attention signals (Google Trends for "SPY options", social media options discussion) could complement equity attention.

\textbf{Cryptocurrencies:} Crypto markets exhibit frequent manipulation due to limited regulation, lower liquidity, and pump-and-dump prevalence ~\citep{hu2021}. AIMM-X could monitor Bitcoin, Ethereum, and altcoins using exchange OHLCV data and crypto-specific attention sources (Crypto Twitter, Discord channels, Telegram groups). Decentralized exchanges and pseudonymous trading complicate detection but also increase manipulator reliance on attention coordination.

\textbf{Fixed income:} Treasury and corporate bond manipulation is less common than equity manipulation but can occur during stress periods or illiquid issues. AIMM-X adaptation requires modeling yield curves, credit spreads, and duration. Attention signals are less obvious (bond markets attract less retail interest), but institutional attention proxies (research downloads, Bloomberg terminal searches) could substitute.

\textbf{International markets:} AIMM-X applies to any market with available OHLCV data and attention proxies. Emerging markets may have higher manipulation prevalence but weaker data infrastructure. Cross-market surveillance detecting coordinated manipulation across geographies (pump U.S.-listed ADRs while manipulating underlying foreign shares) represents an important extension.

\subsection{Ethical Considerations and Societal Implications}

Developing and deploying surveillance tools raises ethical questions requiring careful consideration:

\textbf{False accusation risk:} High integrity scores sometimes reflect legitimate volatility rather than manipulation (META Window 135). Publishing or acting on false accusations causes reputational harm and market disruption. Mitigation requires emphasis on triage rather than definitive claims, transparency about false positive rates, human oversight requirements, and procedural safeguards for accused parties.

\textbf{Privacy considerations:} AIMM-X uses aggregate attention data (subreddit post counts, Wikipedia pageview totals) rather than individual-level information. This design choice prioritizes privacy while enabling detection. Production deployments should maintain this aggregation principle, avoiding surveillance of identifiable individuals unless compelled by legal process.

\textbf{Regulatory capture:} If AIMM-X becomes influential in regulatory enforcement, manipulators could study the methodology and adapt tactics to evade detection. This is the eternal arms race in adversarial domains. Transparency vs. security tradeoffs are real but overstated—most manipulation is unsophisticated, and detection evasion requires resources exceeding most schemes' profitability. Periodic methodology updates (not publicly announced) can mitigate adaptive evasion.

\textbf{Computational inequality:} Advanced surveillance systems (like proprietary platforms) are available only to well-resourced institutions, creating detection inequality where small institutions and retail investors lack protection. AIMM-X partially addresses this through accessibility, though data costs (premium APIs) and expertise requirements remain barriers. Open-source development and hosted deployments could further democratize access.

\textbf{Manipulation vs. coordination:} Not all coordinated trading is manipulation. Reddit communities discussing stocks engage in legal collective speech and investment decisions. Drawing the line between protected activity and illegal coordination is a legal and philosophical challenge beyond AIMM-X's scope. The framework detects anomalies; courts and regulators determine legality.

\textbf{Unintended consequences:} Could widespread AIMM-X deployment reduce market liquidity if attention-coordinated trading (some of which is legitimate) is suppressed? Could false positives discourage retail participation? Could adversarial manipulation increase sophistication in response? These questions require empirical study as deployment expands.

\section{Complete Ticker Statistics}
\label{sec:appendix_ticker_stats}

Table~\ref{tab:full_ticker_stats} provides comprehensive statistics for all 24 tickers.

\begin{table}[h]
\centering
\caption{Complete per-ticker statistics (all windows).}
\label{tab:full_ticker_stats}
\tiny
\begin{tabular}{lrrrrrrrr}
\toprule
Ticker & Windows & Mean $M$ & Median $M$ & Max $M$ & Std $M$ & Mean Dur (bars) & Total Bars & Coverage \% \\
\midrule
META & 12 & 597,231 & 16 & 7,166,596 & 2,068,813 & 15.0 & 36 & 100 \\
MSTR & 10 & 406,687 & 16 & 4,066,778 & 1,286,025 & 16.5 & 33 & 100 \\
SNAP & 8 & 464,895 & 24 & 3,718,879 & 1,314,807 & 23.1 & 37 & 100 \\
AMZN & 11 & 281,293 & 15 & 3,094,132 & 932,913 & 15.5 & 34 & 100 \\
RBLX & 11 & 216,315 & 15 & 2,379,285 & 717,376 & 15.0 & 33 & 100 \\
SPCE & 10 & 174,682 & 22 & 1,746,746 & 552,367 & 22.0 & 44 & 98 \\
NFLX & 11 & 142,377 & 16 & 1,565,992 & 472,160 & 15.9 & 35 & 100 \\
SPY & 8 & 140,994 & 11 & 1,127,905 & 398,772 & 11.3 & 18 & 100 \\
HOOD & 9 & 109,575 & 22 & 986,040 & 328,674 & 22.2 & 40 & 97 \\
TLT & 13 & 71,825 & 14 & 933,662 & 258,950 & 13.8 & 36 & 99 \\
COIN & 7 & 131,899 & 16 & 923,243 & 348,950 & 15.7 & 22 & 100 \\
TSLA & 10 & 87,031 & 15 & 870,173 & 275,168 & 15.0 & 30 & 100 \\
NOK & 10 & 57,396 & 15 & 573,894 & 181,479 & 15.0 & 30 & 95 \\
XLF & 11 & 28,886 & 16 & 317,634 & 95,767 & 15.9 & 35 & 100 \\
IWM & 8 & 14,114 & 14 & 112,808 & 39,879 & 14.4 & 23 & 100 \\
GME & 9 & 31 & 27 & 147 & 45 & 27.2 & 49 & 89 \\
PLTR & 10 & 28 & 20 & 109 & 34 & 20.0 & 40 & 93 \\
AMC & 10 & 16 & 19 & 67 & 20 & 18.5 & 37 & 91 \\
BB & 9 & 23 & 19 & 54 & 17 & 19.4 & 35 & 94 \\
AAPL & 9 & 13 & 21 & 39 & 11 & 20.6 & 37 & 100 \\
GOOG & 11 & 11 & 15 & 25 & 7 & 14.5 & 32 & 100 \\
QQQ & 8 & 10 & 16 & 16 & 5 & 16.3 & 26 & 100 \\
NVDA & 7 & 8 & 18 & 14 & 4 & 17.9 & 25 & 100 \\
MSFT & 11 & 7 & 12 & 12 & 3 & 12.3 & 27 & 100 \\
\bottomrule
\end{tabular}
\end{table}
\FloatBarrier

Note: Coverage\% indicates percentage of trading days with complete OHLCV and attention data.

\section{Temporal Distribution Analysis}
\label{sec:appendix_temporal}

Windows cluster in specific periods during 2024, reflecting known volatility regimes:

\textbf{January (42 windows):} Market reopening volatility, inflation data releases, Fed commentary speculation.

\textbf{March (28 windows):} FOMC meeting cycle, bank stress (SVB anniversary), earnings season.

\textbf{May (19 windows):} Meme stock resurgence, tech earnings, AI hype cycle.

\textbf{August (23 windows):} Summer volatility, yen carry trade unwind, recession fears.

\textbf{November (31 windows):} U.S. election period, policy uncertainty, year-end positioning.

This temporal clustering provides face validity—detected windows align with periods when market participants and media reported elevated volatility and attention.

\section{Data Availability and Reproducibility}
\label{sec:appendix_reproducibility}

\textbf{OHLCV data:} Available from Polygon.io (free tier sufficient for replication, premium tier recommended for extensions). Historical daily bars are permanent and unrevised (split-adjusted).

\textbf{Attention data:} The replication package includes complete proxy generation code for all five attention sources (Reddit, StockTwits, Wikipedia, Google Trends, LIME), detailed documentation of data collection methodology, and configuration files specifying all parameters. Regenerating attention signals requires free-tier API access: Reddit API, Twitter/X API (for StockTwits), Wikipedia API, and Google Trends (public endpoints, subject to rate limiting). The methodology is fully documented, but exact numerical replication may vary slightly due to API changes, rate limits, or retroactive content deletions on source platforms.

\textbf{Code:} Python implementation will be released open-source upon publication with detailed README, example configuration, and step-by-step replication instructions.

\textbf{Computational requirements:} Standard laptop/workstation sufficient (4-core CPU, 16GB RAM). No GPU required. Complete pipeline runtime $< 15$ minutes for 24 tickers with daily data.

\textbf{Output artifacts:} All figures, tables, and window lists reported in this paper are deterministically generated from pipeline and included in replication package.

\textbf{License:} Code released under MIT license. Data subject to provider terms (Polygon.io API terms of service).

\textbf{Contact:} Correspondence to sandeepneela@gmail.com for replication assistance, data requests, or collaboration inquiries.

\section{Glossary of Terms}
\label{sec:appendix_glossary}

\begin{itemize}[leftmargin=*]
\item \textbf{AIMM-X:} AI-driven Market Integrity Monitor with Explainability
\item \textbf{Suspicious Window:} Contiguous time interval where return, volatility, and attention simultaneously deviate from baseline
\item \textbf{Integrity Score ($M$):} Composite score ranking windows by suspiciousness, decomposable into factor contributions
\item \textbf{Phi-signals ($\phi_1$–$\phi_6$):} Interpretable factors measuring return shock, volatility anomaly, attention spike, alignment, recurrence, and disagreement
\item \textbf{Hysteresis Segmentation:} Two-threshold method for window extraction reducing fragmentation
\item \textbf{Attention Fusion:} Weighted combination of multiple attention sources into unified signal
\item \textbf{Z-score:} Standardized deviation measuring how many standard deviations an observation differs from baseline
\item \textbf{OHLCV:} Open, High, Low, Close, Volume—standard price/volume time series
\item \textbf{Triage System:} Detection framework surfacing candidates for human review rather than definitive classification
\item \textbf{Factor Decomposition:} Breaking composite score into constituent components to explain why window was flagged
\end{itemize}

\end{document}